\documentclass{aa}  

\usepackage{graphicx}
\usepackage{txfonts}
\usepackage{color}
\usepackage[normalem]{ulem}

\begin{document}

   \title{Formation of globular clusters with multiple stellar populations
from massive gas clumps in high-z gas-rich dwarf galaxies}
\titlerunning{Formation of globular clusters}


   \author{K. Bekki 
          \inst{1}
          }

   \institute{ ICRAR,
M468,
The University of Western Australia
35 Stirling Highway, Crawley
Western Australia, 6009, Australia \\
              \email{kenji.bekki@uwa.edu.au}
             }

   \date{Received September 15, 1996; accepted March 16, 1997}

 
  \abstract
   {
One of the currently favored scenarios  for the formation of globular clusters (GCs)
with multiple stellar populations is that
an initial  massive stellar system forms (`first generation', FG), subsequently giving rise to gaseous ejecta which is converted into a second generation (SG) of stars
to form a GC.
How such GCs with such FG and SG populations
form and evolve, however, remains unclear.
   }
   {
We therefore investigate, for the first time,
the sequential formation processes of both FG and SG stars
from star-forming massive gas clumps
in gas-rich dwarf disk galaxies.
  }
{

We adopt a novel approach to resolve 
the two-stage formation of GCs in hydrodynamical simulations of dwarf galaxies.
In the new simulations, new gas particles
that are much less massive than their parent star particle  are generated around
each new star particle  when the new star enters into  the asymptotic giant branch
(AGB) phase.
Furthermore, much finer maximum time step width ($\sim 10^5$ yr) 
and smaller softening length ($\sim 2$ pc)
are adopted for such AGB gas particles to properly resolve the ejection
of gas from AGB stars and AGB feedback effects.
Therefore, secondary star formation from AGB ejecta 
can be properly investigated
in galaxy-scale simulations.
}
{
An FG stellar system can first form from a massive gas clump
developing due to gravitational instability within its host gas-rich dwarf galaxy.
Initially the FG stellar system is not a single massive cluster, but
instead is composed of several
irregular stellar clumps (or filaments)
with a total mass larger than $10^6 M_{\odot}$.
While the FG system is dynamically relaxing,
gaseous ejecta from AGB
stars can be gravitationally 
trapped by the FG system
and subsequently converted into new stars to form a compact SG stellar system
within the FG system.
Interestingly, about 40\% of AGB ejecta is from stars that do not
belong to the FG system (`external gas accretion').
FG and SG stellar systems have
different amplitudes of internal rotation and $V/\sigma$.
The mass-density ($M_{\rm SG}-\rho_{\rm SG}$)
relation for  SG stellar systems can be approximated as
$\rho_{\rm SG} \propto M_{\rm SG}^{1.5}$.
There can be a threshold total mass of GC host galaxies
($M_{\rm th} =[5-23]\times {10}^9  M_{\odot}$)
beyond which the formation of GCs with compact SG stellar systems
is possible.
Both
the initial baryonic mass fraction 
and the gas mass fraction in  dwarfs are crucial parameters
that determine whether or not  GCs can contain multiple stellar populations. 
GCs with compact SG stellar systems are more likely to form
in dwarf disks with larger  gas mass fractions and higher surface mass densities.
Formation of binary GCs with SGs and the subsequent GC  merging are clearly seen in some
models. The derived external gas-accretion process in FG systems initially  consisting
of stellar clumps
will need to be investigated further in more sophisticated simulations.
}
   {}


   \maketitle
%

\section{Introduction}

A growing number of recent photometric and spectroscopic observations of the Galactic
globular clusters (GCs) reveal strong evidence
for the presence of multiple stellar populations in GCs (e.g.,
Bedin et al. 2004; Piotto et al. 2005, 2007; Milone et al. 2010;
Gratton et al. 2013; see Gratton, Carretta, and Bragaglia 2012, GCB12, for a recent review).
One key recent observation
is that the majority ($\sim 70$\%) of stellar populations
in GCs are so-called `second generations' (SGs) of stars which form from gas ejecta
from
stars of an earlier generation (often referred to as first generation (FG) stars) (e.g., Carretta et al. 2010a).
The observed large fraction of SG stars and Na-O and C-N
 anti-correlations between these stars
have had a significant impact on theoretical studies of GC formation and have accordingly  prompted
a more sophisticated and realistic  modeling of GC formation (e.g., D'Ercole et al. 2008; D08).
Recent observational studies on
possible differences in  physical properties (e.g., radial distributions and kinematics)
 between FG and SG  stars
have provided new valuable constraints on the formation processes of GCs with multiple stellar
populations
(e.g., Sollima et al. 2007; Bellini et al. 2009;
D'Orazi et al. 2010; Mackey et al. 2013; Larson et al. 2013).

The degrees of abundance inhomogeneity in GCs with multiple stellar populations
are observed to be diverse (GCB12; Marino et al. 2011, 2012, and 2015;
Johnson et al. 2015).
At least eight Galactic GCs have been demonstrated to have metallicity spread
(not just abundance spread in light elements), and some of them show abundance
spread even in $s$-process elements (e.g., Marino et al. 2015), which implies that
both Type II supernovae (SNe) and asymptotic giant branch (AGB) stars might have possibly polluted the early
chemical evolution of GCs. The two metallicity groups in M22 have significantly different
[Ba/Fe], [Y/Fe], [La/Fe], and [La/Eu] (e.g., Marino et al. 2011),
which provides strong constraints both on the $s$-process enrichment process
of forming GCs and on the chemical yields of AGB stars (e.g., Shingles et al. 2014).
Carretta (2015) recently discovered five distinct stellar populations in NGC 2808,
which implies that there are more than just two major star formation epochs
,  as was modeled in
previous theoretical studies of GCs with multiple stellar populations (e.g., D08).
The observed variety of multiple stellar populations in the Galactic GCs needs
to be explained by a theoretical model of GC formation.

One of the possible GC formation scenarios in recent observational and theoretical
works related to multiple stellar populations
of GCs  implies that the original GCs were much more massive
than the present-day ones (e.g., Renzini 2015).
In this scenario,  massive stellar systems with total masses
approximately 10 times larger than the present typical GC mass of
$2 \times 10^5 M_{\odot}$ were first formed from gas clouds.
Numerous massive rotating stars (e.g., Decressin et al. 2008; Krause et al. 2013) or AGB stars
(e.g., D'Antona \& Caloi 2004; D08) in the FG stellar systems
ejected gas from which SG stars were formed.
The main observational fact  on which this `two-stage' GC formation scenario is based is
that the mass fractions of SG stars in the
Galactic  GCs are  as large as $50-80$\% (e.g., D'Antona \& Caloi 2008; Carretta et al. 2009; GCB12).
It is still unclear in this scenario whether it is gaseous ejecta
from massive rotating or AGB stars
that is  responsible for the formation of SG stars in GCs.
Bastian et al. (2013) recently proposed that accretion of gas from
interacting massive binary and rapidly rotating stars onto circumstellar disks
of low-mass pre-main sequence stars is responsible for the origin of multiple
stellar populations of GCs. In this scenario, there is no age difference
(thus no `FG-SG dichotomy')
between multiple stellar populations of GCs, which is quite different from
other models such as D08.

Previous numerical simulations of GC formation demonstrated that SG stars can form from
AGB ejecta of FG stars in the central regions of  forming GCs (D08; Bekki 2011; B11).
These previous theoretical works
mainly investigated SG formation from ejecta of FG stars
{\it in  fixed or live gravitational potentials of already existing FG stellar systems} 
in GCs.
Recent simulations investigated SG formation in molecular clouds with fractal
structures; however,  they did not include galaxy-scale hydrodynamics in a
self-consistent manner (Bekki 2017a, b).
Furthermore, the mixing of AGB ejecta from 
FG stars  and pristine gas from their host galaxy,
which is essential for the observed Na-O anti-correlation
(e.g., D08),
was ignored in these previous simulations.
Therefore, it remains unclear (i) how FG stars 
form and evolve at the epoch of GC formation 
within their host galaxies
and (ii) how AGB ejecta can be mixed with cold interstellar
medium of their host dwarfs  in these
previous works.
Dynamical evolution of GCs with both FG and SG stars with initially different spatial
distributions and kinematics have just recently begun to be investigated (e.g., Vesperini et al. 2010; 2013),
and the results have important implications both for the long time evolution of GCs
and for the origin of the Galactic stellar halo (e.g., Vesperini et al. 2010).
It is thus important for theoretical studies of GC formation to predict initial
dynamical properties of GCs.

Although previous galaxy-scale and cosmological simulations have tried to identify the possible
formation sites of GCs in galaxies (e.g., Bekki \& Couch 2001; Bekki et al. 2002;
Bromm \& Clarke 2002;
 Kravtsov \& Gnedin 2005; Saitoh et al. 2011;
Kruijssen et al. 2012;
Renaud et al. 2015),  the SG star content of the identified `GC' candidates was not investigated: they may or may not be genuine
GCs. Semi-analytic models of galaxy formation based on a cold dark matter (CDM) cosmology
assumed GC formation in galactic building blocks (e.g., dwarf galaxies) at high redshifts in order to
investigate the origin of physical properties of GC systems in galaxies
(e.g., Beasley et al. 2002;  Bekki et al. 2008; Griffen et al. 2010; Tonini 2013).
Elmegreen et al. (2012) have recently proposed that high-redshift dwarf galaxies with strong
Ly$\alpha$ emission ($\ge 10^{42}$ erg s$^{-1}$) are the formation sites of metal-poor GCs.
It is not understood, however, in what physical conditions genuine GCs with both FG and SG stars
can be formed in galactic building blocks at high redshifts  in these studies.
Therefore, it remains theoretically unclear 
(i) how FG stars were formed and evolved within forming GCs
and (ii) whether or not GCs with multiple stellar populations
can really be formed within galaxies; see Forbes et al. (2018)
for more details of the open
questions related to  GC formation in the early universe.

The purpose of this paper is therefore to investigate the formation processes of
{\it both FG and SG stars} in  GCs
by using  self-consistent numerical simulations of GC formation.
We develop our new simulation code that is specially purposed to investigate star formation from
gas ejected from AGB stars in  FG stellar systems of  forming GCs. By using this new code,
we investigate the following five points in particular: (i)  How FG stellar systems
can be formed in the early phase of GC formation, (ii) whether and in what physical conditions
compact SG stellar systems can
be formed during GC formation, (iii) what different physical properties
FG and SG stellar systems have that can be observed,
(iv) what roles of AGB stars and Type II supernovae (SNII) can play
in the formation of FG and SG stellar systems,
and (v) what physical properties of a galaxy  are required
for the galaxy to host GCs with SG stars.
Since our new code does not include full chemical yields both from SNe and AGB stars
self-consistently,
we do not discuss the origin of the observed Na-O and Mg-Al anti-correlations between
cluster stars (e.g., Carretta et al. 2009; 2010a) in the present study.

In these investigation,
we assume that most GCs with multiple stellar populations
were initially formed in gas-rich,
actively star-forming dwarf galaxies at high redshifts,
which have been considered to be the building blocks of luminous galaxies like our Milky Way
(e.g., Searle \& Zinn 1978).
This assumption is fairly reasonable and realistic for the following reasons.
First, some dwarf galaxies in the Local Group  and nearby galaxy groups are observed to have
old and relatively young GCs (e.g., van den Bergh 2000; Georgiev et al. 2009).
Second, the Large Magellanic Cloud (LMC), which is classified as a dwarf irregular galaxy
is observed to show possible evidence of multiple stellar populations,
such as extended main sequence turn-off morphologies,
age spreads, and the presence of young stellar objects within clusters
(e.g., Mackey et al. 2008; Mucciarelli et al. 2009; 
Keller et al. 2012; Li et al. 2016; For \& Bekki 2017, FB17; Milone et al. 2017).
Third, previous and recent semi-analytic models based on a CDM cosmology
assumed that  metal-poor GCs can be formed in massive dwarf galaxies in order to explain
the key physical properties of galactic GC systems such as metallicity distribution functions
of GCs and correlations between GC host luminosities and mean GC metallicities
(e.g., Beasley et al. 2002).
We thus extensively investigate GC formation processes within dwarf galaxies with different
total masses, gas mass fractions, and baryonic mass fractions to clarify the physical conditions
required for GC formation.

Since this paper focuses exclusively
on the entire formation process of FG and SG stars in GCs,
it does not discuss other key issues related to the origin of multiple stellar populations in GCs,
such as GC formation from nucleated dwarfs (e.g., Freeman 1993; Bekki \& Freeman 2003; B\"oker 2008),
chemical evolution of forming GCs (Bekki et al. 2007; D'Ercole et al. 2010),
the importance  of lithium production of AGB stars in the origin of multiple stellar populations
(e.g.,  Ventura \& D'Antona et al. 2010),
the origin of the Galactic GCs with unique characteristics of multiple stellar populations
such as  NGC 1851 (e.g., Yong et al. 2009), NGC 2419 (e.g., Cohen \& Kirby 2012),
and NGC 2808 (e.g., Bragaglia et al. 2010),
the formation of $\omega$ Centauri with
age and  metallicity spreads (e.g., Lee et al. 1999; Hilker et al. 2004; Bellini et al. 2009;
Johnson \& Pilachowski 2010; Marino et al. 2012),
and the origin of He-rich populations in GCs (e.g., Norris 2004; Lee et al. 2005; Piotto et al. 2005),
and GC-halo connections (e.g., Vesperini et al. 2010; Martell et al. 2011).
Each of these key issues would need to be discussed in detail in a separate paper.

The plan of this  paper is as follows: In the following section,
we describe the methods and techniques of
our galaxy-sale simulations with a special model for star formation from AGB ejecta.
In \S 3, we present the numerical results
on the formation processes of both FG and SG stars
and their dependencies on physical properties of GC host dwarfs.
In \S 4, we discuss the important implications of the present results in terms of
(i) physical mechanisms of dilution of AGB ejecta by pristine gas
and (ii) possible internal [Fe/H] spread in stars of individual GCs.
We summarize our  conclusions in \S 5.

The present simulations are different from our previous ones
which could combine direct N-body simulations of star clusters  based on the NBODY code
with hydrodynamical simulations (Hurley \& Bekki 2008). Therefore, we cannot
discuss the long-term dynamical ($>1$ Gyr) evolution of forming clusters within
dwarfs.
We will discuss th important issues in our forthcoming paper.
Fast- rotating massive stars
(FRMS), AGB stars, massive interacting binaries (MIB),
and supermassive stars are suggested to be `polluting' stars, the ejecta of which
can be mixed with pristine ISM and consequently converted into new stars
(e.g., Karakas et al. 2006; Bastian et al. 2013;
Krause tal. 2013; Denissenkov \& Hartwick 2014;
See Renzini 2015 for a critical review for advantages and disadvantages
of each of these polluters in the formation of GCs). We do not discuss
which of the four polluters is the most promising  one for
self-consistently explaining the observed properties of
GCs with multiple stellar populations in this paper.
There are many observational papers on the internal variations of chemical
abundances (e.g., light elements, $s$-process elements,[Ca/H], and [Fe/H])
in GCs with luminosities and  sizes (e.g., Cohen 1981;
Cottrell \& Da Costa 1981; Smith 1987; Kraft 194;
Norris \& Da Costa 1995; Yong et al. 2015).
The origin of internal abundance variations
in each of these individual GCs will need
to be discussed in our forthcoming papers.

   \begin{figure*}
   \centering
   \includegraphics{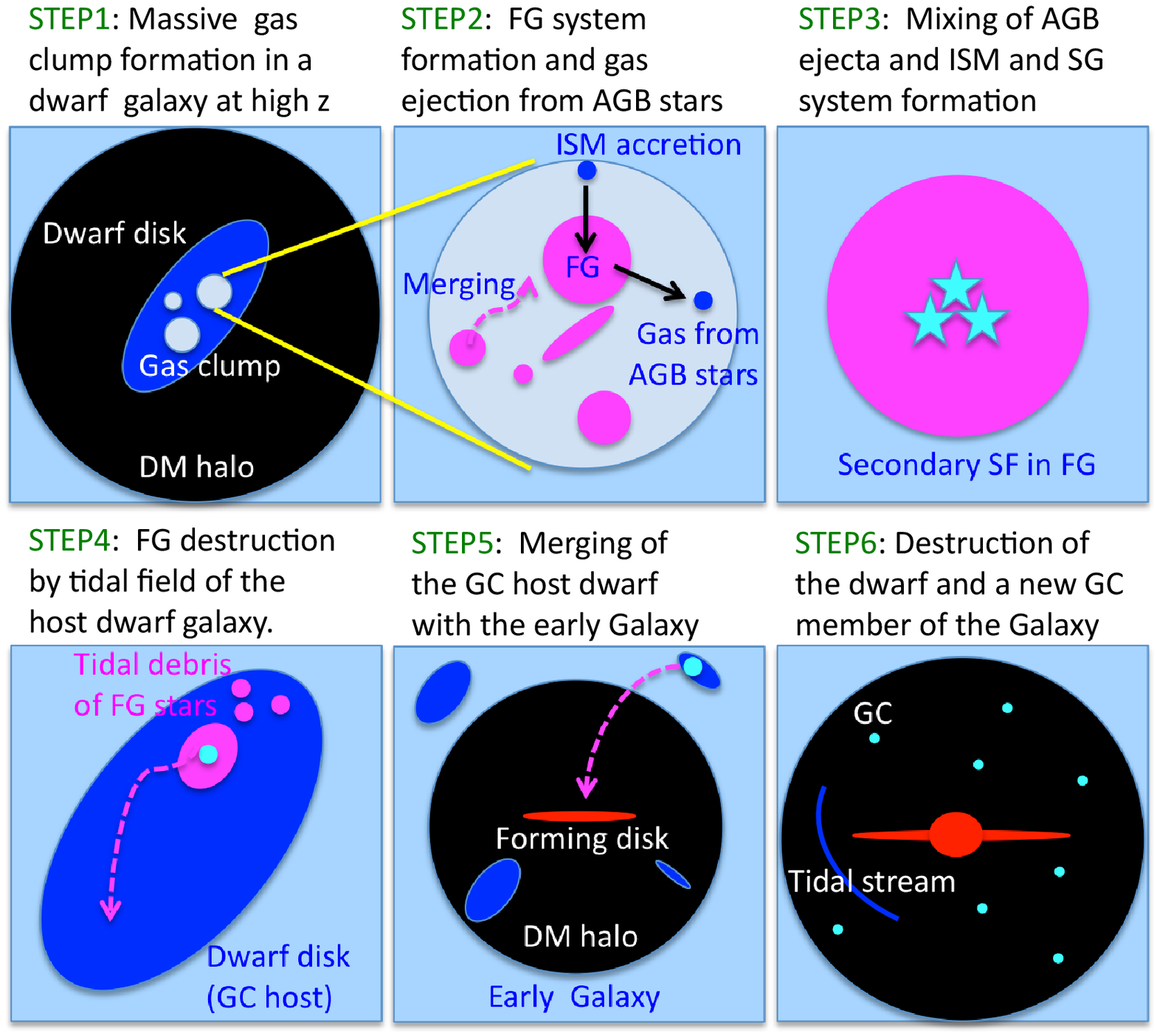}
   \caption{
Formation of Galactic GCs with
multiple stellar populations in a very gas-rich dwarf at high z.
The entire formation process is divided into six steps.
STEP 1: the formation of massive clumps consisting of gas and new
stars in very gas-rich massive dwarf disk galaxies.
STEP 2:
the formation of  massive FG systems  from merging stellar
clumps and filaments  developed in these clumps.
STEP 3: the formation of SG stars from AGB ejecta
mixed with pristine ISM in forming FG systems.
STEP 4: the almost complete
destruction of FG stellar systems by the tidal field of
GC-host dwarfs.
STEP 5: the accretion of GC host dwarfs onto the halo region of the Galaxy
in the early formation history of the Galaxy.
STEP 6: the destruction of GC host dwarfs by the strong tidal field of
the Galaxy.
               }
              \label{FigGam}%
    \end{figure*}

   \begin{table}
      \caption[]{
Description of the basic parameter values
for the standard  model M1.
}
         \label{KapSou}
     $$
         \begin{array}{p{0.5\linewidth}l}
            \hline
            \noalign{\smallskip}
            Physical properties      &  {\rm Parameter} \hspace{2mm} {\rm  values} \\
            \noalign{\smallskip}
            \hline
            \noalign{\smallskip}
Total Mass
& M_{\rm t}=2.25 \times 10^{10}  M_{\odot}  \\
DM mass
&   M_{\rm dm}=2\times 10^{10} M_{\odot} \\
DM core radius
&  a_{\rm dm}=2.8kpc \\
Gas fraction
&   f_{\rm g}=0.8\\
Baryonic mass fraction
&   f_{\rm b}=0.11\\
Stellar disk size
&   2.3 kpc \\
Gas disk size
& 4.5 kpc \\
Initial gaseous Q
& Q_{\rm g}=0.5  \\
Initial stellar  Q
&   Q_{\rm s}=1.5 \\
Initial gas temperature
&  300 K \\
AGB yield
&   VG97\\
Feedback effects
&  SNII and AGB \\
Initial metallicity
& {\rm [Fe/H]_0}=-1.45  \\
Star formation (FG and SG)
&   \rho_{\rm th}=100 \hspace{2mm} atom \hspace{2mm} cm^{-3}  \\
Initial particle number
&  N=10^6 \\
AGB ejecta per particle
&  n_{\rm agb}=1 \\
IMF
& Salpeter \hspace{2mm}  (\alpha_{\rm IMF}=2.35) \\
Softening length (DM)
&  \epsilon_{\rm g,dm}=193 pc\\
Softening length (stars)
&  \epsilon_{\rm g, os}=21 pc \\
Softening length (SG stars)
&  \epsilon_{\rm g, ns}=2 pc\\
Maximum time step width
&  \delta t_{\rm agb}=8.8 \times 10^4 yr.\\
            \noalign{\smallskip}
            \hline
         \end{array}
     $$
   \end{table}

   \begin{table*}
      \caption[]{Summary of model parameters.}
         \label{KapSou}
     $$ 
         \begin{array}{cccccccl}
            \hline
            \noalign{\smallskip}
Model &
M_{\rm dm} (10^9 M_{\odot})&
M_{\rm s}  (10^9 M_{\odot})&
M_{\rm g} (10^9 M_{\odot})&
f_{\rm b} &
f_{\rm g} &
R_{\rm g} (kpc)&
Comments \\
            \noalign{\smallskip}
            \hline
            \noalign{\smallskip}
M1 & 20.0 & 0.5 & 2.0 & 0.11 & 0.80 & 4.5 & standard \hspace{1mm} model \\
M2 & 20.0 & 0.5 & 2.0 & 0.11 & 0.80 & 4.5 & strong SN \hspace{1mm} feedback \\
M3 & 20.0 & 0.5 & 2.0 & 0.11 & 0.80 & 4.5 & no \hspace{1mm} SN \hspace{1mm} feedback \\
M4 & 20.0 & 0.5 & 2.0 & 0.11 & 0.80 & 4.5 & no \hspace{1mm} AGB \hspace{1mm} feedback \\
M5 & 20.0 & 0.5 & 1.0 & 0.07 & 0.67 & 4.5 &  \\
M6 & 20.0 & 0.5 & 1.5 & 0.09 & 0.75 & 4.5 &  \\
M7 & 20.0 & 1.5 & 1.0 & 0.11 & 0.40 & 4.5 &  \\
M8 & 20.0 & 2.0 & 0.5 & 0.11 & 0.20 & 4.5 &  \\
M9 & 20.0 & 0.1 & 0.4 & 0.02 & 0.80 & 4.5 &  \\
M10 & 20.0 & 0.1 & 0.6 & 0.03 & 0.86 & 4.5 &  \\
M11 & 20.0 & 0.1 & 1.0 & 0.05 & 0.91 & 4.5 &  \\
M12 & 20.0 & 0.1 & 2.0 & 0.10 & 0.95 & 4.5 &  \\
M13 & 20.0 & 0.5 & 2.0 & 0.11 & 0.80 & 11.3 &  LSB\\
M14 & 20.0 & 0.5 & 2.0 & 0.11 & 0.80 & 7.1 &  LSB\\
M15 & 20.0 & 0.5 & 2.0 & 0.11 & 0.80 & 2.3 &  higher \hspace{1mm} surface 
\hspace{1mm}  density \\
M16 & 2.0 & 0.05 & 0.2 & 0.11 & 0.80 & 1.4 &   \\
M17 & 6.0 & 0.15 & 0.6 & 0.11 & 0.80 & 2.5 &   \\
M18 & 60.0 & 1.5 & 6.0 & 0.11 & 0.80 & 7.8 &   \\
M19 & 2.0 & 0.05 & 0.2 & 0.11 & 0.80 & 0.7 &  higher \hspace{1mm} surface \hspace{1mm} density \\
M20 & 6.0 & 0.15 & 0.6 & 0.11 & 0.80 & 1.2 &  higherpace{1mm} surface \hspace{1mm}  density \\
M21 & 60.0 & 1.5 & 3.0 & 0.07 & 0.67 & 7.8 &   \\
M22 & 60.0 & 1.5 & 1.5 & 0.05 & 0.50 & 7.8 &   \\
M23 & 20.0 & 0.5 & 2.0 & 0.11 & 0.80 & 4.5 & T_{\rm g}=10^3K \\
M24 & 20.0 & 0.5 & 2.0 & 0.11 & 0.80 & 4.5 & T_{\rm g}=10^4K \\
M25 & 20.0 & 0.5 & 2.0 & 0.11 & 0.80 & 4.5 & Q_{\rm g}=0, Q_{\rm s}=1.5 \\
M26 & 20.0 & 0.5 & 2.0 & 0.11 & 0.80 & 4.5 & Q_{\rm g}=Q_{\rm s}=1.5 \\
M27 & 20.0 & 0.5 & 2.0 & 0.11 & 0.80 & 4.5 & Q_{\rm g}=Q_{\rm s}=3.0 \\
M28 & 20.0 & 0.5 & 2.0 & 0.11 & 0.80 & 4.5 & \rho_{\rm th, SG}=1 \hspace{1mm} atom \hspace{1mm} cm^{-3}\\
M29 & 20.0 & 0.5 & 2.0 & 0.11 & 0.80 & 4.5 & \rho_{\rm th, SG}=1000 \hspace{1mm} atom \hspace{1mm} cm^{-3}\\
            \noalign{\smallskip}
            \hline
         \end{array}
     $$ 
\end{table*}

\section{The model}

\subsection{A possible scenario}

Figure 1 briefly illustrates a GC formation scenario, of which the details are investigated and discussed in this paper. The entire GC formation
process from GC host galaxy formation to GC migration into the Galactic
halo is divided into six physical processes 
(STEP 1 - 6) for convenience in Figure 1.
STEP 1 is the formation of massive clumps consisting of gas and new
stars in very gas-rich massive dwarf disk galaxies that can
be ubiquitous at high $z$.  STEP 2 is
the formation of  massive FG systems  from merging stellar 
clumps and filaments  developed in these clumps.
As shown later in this paper,
the formation process of FG stellar systems is rather complicated,
which influences the formation of SG stars within FG systems.
Therefore,  the formation
of massive clumps is a key physical process of FG system formation,
and a better understanding of the physical conditions required for
the clump formation will hopefully lead to a comprehensive understanding of
why GCs with multiple stellar populations can be formed mostly 
at high z. The minimum halo mass 
($M_{\rm h, min}$) required for a dwarf to host a GC is described
as follows:

\begin{equation}
M_{\rm h, min}= 4.6 \times 10^9 M_{\odot} ( \frac{ 1+F_{\rm g} }{3} )
( \frac{ S_{\rm N} }{5} )^{-1} ( \frac{ F_{\rm b} }{0.1} )^{-1},
\end{equation}
where $F_{\rm g}$ is the mass ratio of cold gas to stars,
$S_{\rm N}$ is the specific frequency of GCs (e.g., Harris \& van den Bergh 1981), and
$F_{\rm b}$ is the mass ratio of baryonic components (cold gas and stars)
to dark matter (later we use $f_{\rm b}$ for simulations to distinguish
between observed and simulated baryonic fractions).
Since higher $S_{\rm N}$ and $F_{\rm b}$ are adopted
as reference values above, $M_{\rm h, min}$ can be significantly higher
than the above in real galaxies.
For example, 
$M_{\rm h, min}= 2.3 \times 10^{10} M_{\odot}$
for  $S_{\rm N}=2$ and $F_{\rm b}=0.05$.
$M_{\rm h, min}$ in the above equation (1) is a reasonable guideline
to simulate GC formation in dwarfs.

STEP 3 is the formation of SG stars in forming FG systems: gas ejection
from AGB stars, accretion of the ejecta onto the FG systems,
and conversion of the ejecta. The present study focuses exclusively
on the formation of new stars from AGB ejecta, and does not discuss
the importance of other polluters (e.g., FRMS and massive binary stars).
STEP 2 and 3 correspond to the `two-stage' formation process of GCs with
multiple stellar populations.
STEP 4 is the almost complete
destruction of FG stellar systems by the tidal field of
GC-host dwarfs.  Other mechanisms of  FG destruction,
such as expansion through gas expulsion  (e.g., Khalaj \& Baumgardt 2016),
could also be  possible.
STEP 5 is the accretion of GC host dwarfs onto the halo region of the Galaxy
in the early formation history of the Galaxy.
STEP 6 is the destruction of GC host dwarfs by the strong tidal field of
the Galaxy.  During this tidal destruction,  GCs within the dwarfs can
be stripped to become one of the Galactic halo GCs.
These accretion events of dwarfs with GCs are key physical processes
in interpreting the observational data of GC properties in the Galactic halo
(e.g., Forbes \& Bridges 2010).
STEP 4- 6 are  not investigated
in the present study, and will be discussed in our forthcoming papers
based on separate numerical simulations on dynamical evolution of GCs
and dwarfs.

Formation of massive gas clumps in relatively gas-rich luminous disk
galaxies was already investigated by Shlosman \& Noguchi (1993)
and the importance
 of such gas clumps in various galaxy formation processes
was also discussed by Noguchi (1999).
Bekki (2007) also showed that massive clumps consisting of
gas and stars can be formed in gas-rich dwarf galaxies and discussed
the formation of stellar galactic nuclei through merging of such GC-like clumps.
However, these previous simulations did not discuss the roles
of massive clumps in the formation of GCs with multiple stellar populations.
The present study therefore investigates whether and how massive 
gaseous and stellar clumps (corresponding to FG stellar systems) can be
formed in gas-rich dwarf disk galaxies in detail.

Although GC formation in dwarf galaxies has been modeled 
in previous theoretical models 
(e.g., Beasley et al. 2002;
Bromm \& Clarke 2002;
Kravtsov \& Gnedin 2005; Bekki et al. 2008),
whether  `GCs' have multiple stellar populations has not been discussed.
Currently, it is well known that almost all old GCs in the Galaxy contain
multiple stellar populations (Carretta et al. 2009). Therefore,
it is not clear whether or not `GCs' in these previous simulations are genuine
GCs.
Although recent simulations of cluster formation by Renaud et al. (2015)
discussed prolonged star formation within clusters, they did not include
AGB ejecta in their simulations. Therefore, their models did not enable the
authors to discuss whether the simulated clusters can become GCs with
internal abundance spreads in light elements.
The present study, for the first time, tries to select `genuine' GCs with
at least two stellar populations in the hydrodynamical simulations of
dwarf galaxy evolution with secondary star formation from AGB ejecta. 

   \begin{figure*}
   \centering
   \includegraphics{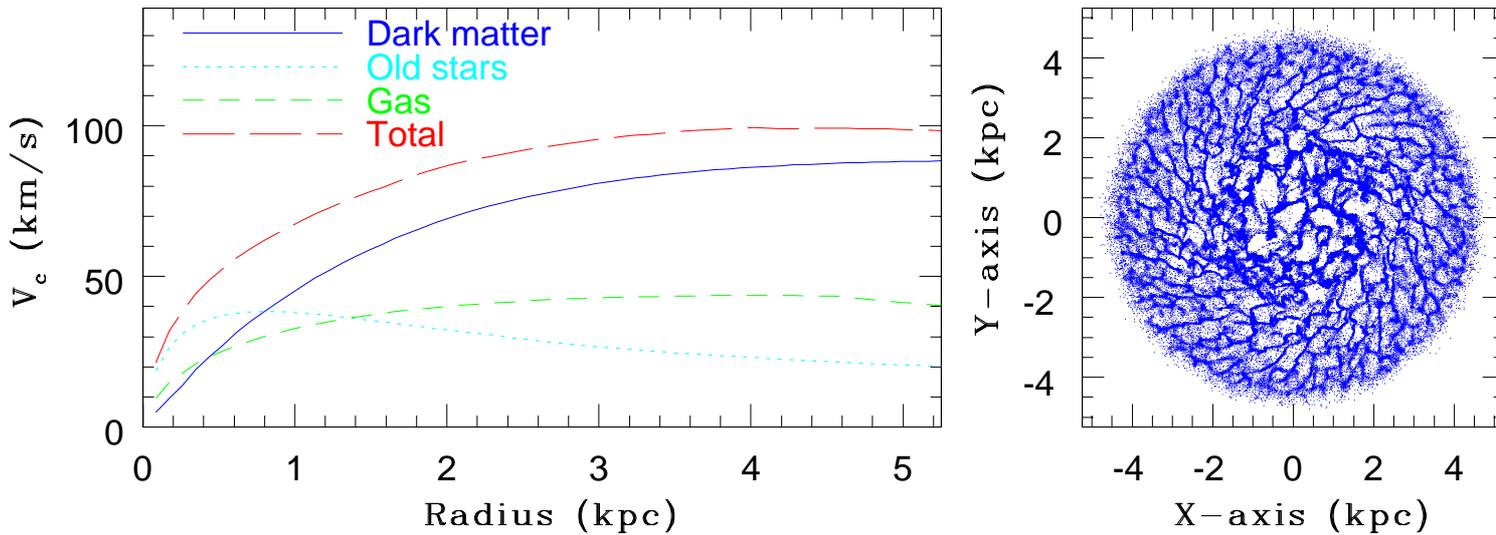}
   \caption{
The initial contributions of dark matter,  old stars,  and gas to
the rotation curve of a gas-rich dwarf at $T=0$ Myr
and the gas distribution projected onto the $x$-$y$ plane
at $T=56$ Myr
in the standard model (M1).
}
              \label{FigGam}%
    \end{figure*}

\subsection{The required large baryonic fraction for GC formation}

In the above equation (1),  a reference value of $F_{\rm b}=0.1$ is adopted
to discuss a plausible $M_{\rm h, min}$: it should be stressed
here that $F_{\rm b}$ is the mass fraction of gas and stars.
As described later in this paper
(Section 3),
such a high $F_{\rm b}$ ($>0.05$)  is indeed required for
GC formation within
the simulated dwarf galaxies.
However, it should be noted that the required high $F_{\rm b}$ would be
much larger than the average of $F_{\rm b}$ 
for $M_{\rm h} \sim 10^{10} M_{\odot}$ in recent theoretical studies
of galaxy formation based on $\Lambda$CDM models (e.g., Moster et al. 2013).
The results of the  semi-analytic model of galaxy formation presented by these latter authors
accordingly imply 
that only rare dwarf galaxies with very high $F_{\rm b}$ can
form genuine GCs with FG and SG stars in the present scenario.
Moster et al. (2013) also showed that more massive halos can have higher 
$F_{\rm b}$ for $M_{\rm h} < 10^{12} M_{\odot}$. 
Their results suggest that only high-mass halos with $M_{\rm h} > 10^{11}
M_{\odot}$ can have the required high $F_{\rm b}$ for GC formation.
These theoretical results
imply that the required high $F_{\rm b}$ in the simulated galaxies
of the present study
could be a potentially serious problem in the present GC formation
scenario.
It should be noted, however, that  there are a number of observed galaxies with 
$M_{\rm h}=10^{10}-10^{11} M_{\odot}$  that have relatively high $F_{\rm b}$
(e.g., Fig. 16 Papastergis et al. 2012). 
The Small Magellanic Cloud is also an  example galaxy where
the baryonic mass fraction is rather large (Bekki \& Stanimirovic 2009).
Therefore,  the present models with high $F_{\rm b}$ are not completely
inconsistent with observations.
It may also be possible that high-z dwarfs have higher $F_{\rm b}$ 
within their disks than the low-z counterparts.

The possible threshold halo mass for GC formation (described later in \S 3)
implies that the typical metallicity of GCs can be high.
We can discuss this point briefly, using either the observed
mass-metallicity relation of dwarfs or the theoretical prediction
from semi-analytic models of galaxy formation (e.g., Guo et al. 2016).
Gas-rich dwarf galaxies with $M_{\rm h}=5 \times 10^{9} M_{\odot}$
and $M_{\rm s}=10^8 M_{\odot}$ can have a metallicity
of [Fe/H]$\sim -1$, according to the mass-metallicity relation
($Z \sim M^{0.46}$)
derived by Tremonti et al. (2004). 
Accordingly, the present GC formation scenario,
in which $M_{\rm h, min} \sim 5 \times 10^9 M_{\odot}$,
suggests that most GCs can have [Fe/H]$\sim -1$. This is not consistent
with the observed metallicity distribution function of the Galactic GCs,
which shows a metallicity peak around [Fe/H]$=-1.6$ (e.g., Harris 1999).
However, the mass-metallicity (or luminosity-metallicity) relation at high $z$
could be quite different from that at $z=0$
(e.g., Guo et al. 2016). Furthermore, there is a large scatter in
metallicities of galaxies for a given galaxy mass
at higher redshifts in theoretical predictions 
(e.g., Guo et al. 2016).
Also, $M_{\rm s}$ can be
significantly lower than $10^8 M_{\odot}$ (thus lower
metallicities) in the halos with
$M_{\rm h}=5 \times 10^9 M_{\odot}$ at high $z$.
We therefore do \textcolor[rgb]{0.984314,0.00392157,0.0235294}{not} consider that the observed typical metallicity of GCs
is \textcolor[rgb]{0.984314,0.00392157,0.0235294}{not}
a problem for the present scenario of GC formation: if the GC-host
dwarfs have metallicities slightly smaller than the  observed ones
for their masses
(e.g., Tremonti et al. 2004), then the typical metallicity of GC can
be reproduced.
Furthermore, as described later, the present results do not depend
on the adopted [Fe/H].

\subsection{GC host dwarfs}

In order to perform  numerical simulations of  GC formation in dwarf disk
galaxies on GPU clusters,
we have revised our previous code (`GRAPE-SPH'; Bekki 2009),
which can be run on the special computer for gravitational dynamics (GRavity PipE;
Sugimoto et al. 1990).
In the present paper, we describe only the key ingredients of the older version of the code
and focus on the new physics that is incorporated into the revised version
(e.g., inclusion of star formation from AGB ejecta).
Since we mainly investigate dynamical processes of GC formation,
we do not include chemical evolution in the present simulations in a fully self-consistent manner.

A  dwarf disk galaxy is modeled as a fully self-gravitating system
and is assumed to consist of a dark matter halo and stellar and gas disks.
The dark matter halo and the main stellar component of the dwarf
are represented by collisionless N-body particles whereas
the gas component is represented by 
SPH particles.
The total masses of the dark matter halo, the stellar disk, and the gaseous disk in the dwarf
are represented by $M_{\rm dm}$, $M_{\rm s}$, and $M_{\rm g}$, respectively.
The baryonic mass fraction ($f_{\rm b}=(M_{\rm s}+M_{\rm g})/M_{\rm t}$,
where $M_{\rm t}=M_{\rm dm}+M_{\rm s}+M_{\rm g}$) and 
gas mass fraction ($f_{\rm g}=M_{\rm g}/(M_{\rm s}+M_{\rm g}$)) 
are key parameters that can determine whether GCs with FG and SG stars can be
formed in dwarfs.

The density profile of the dark matter halo
is represented by that proposed by
Salucci \& Burkert (2000):
\begin{equation}
{\rho}_{\rm dm}(r)=\frac{\rho_{\rm dm,0}}{(r+a_{\rm dm})(r^2+{a_{\rm dm}}^2)},
\end{equation}
where $\rho_{\rm dm,0}$ and $a_{\rm dm}$ are the central dark matter
density and the core (scale) radius, respectively.
For convenience, we hereafter refer to this profile as the ``SB'' profile (or
model).
Recent observational and numerical studies have shown that the adopted ``cored
dark matter'' halos are reasonable for describing dark matter distributions
in low-mass galaxies (e.g., Gavernato et al. 2010; Oh et al. 2011).
Therefore, the above SB profile rather than the ``NFW'' one (Navarro et al. 1996)
with a central cusp predicted by the cold dark matter (CDM)  model
is better for the present model for  dwarfs.
For the SB profile, the dark matter
core parameters, $\rho_{\rm dm,0}$,  $a_{\rm dm}$,  and $M_{0}$
(where $M_{0}$ is the total dark matter mass within $a_{\rm dm}$),
are not free parameters, and clear correlations are observed between
them (Burkert 1995):
\begin{equation}
M_{0}=C_0 {(\frac{a_{\rm dm}}{\rm kpc})}^{7/3} M_{\odot},
\end{equation}
where $C_0$ is $4.3 \times 10^7$ in the original formula by Burkert (1995).
We mainly present the results of the models with 
$C_0=2.2 \times 10^7$ (i.e., more compact dark matter halo).
All dark matter particles are distributed within 5$a_{\rm dm}$.

The stellar component of the dwarf is modeled as a  bulge-less stellar disk
with the size of $R_{\rm s}$.
The radial ($R$) and vertical ($Z$) density profiles of the stellar disk are
assumed to be proportional to $\exp (-R/R_{\rm s,0}) $ with scale
length $R_{\rm s,0} = 0.2R_{\rm s}$ and to ${\rm sech}^2 (Z/Z_{\rm s,0})$ with scale
length $Z_{\rm s,0} = 0.04R_{\rm s}$, respectively.
In addition to the
rotational velocity caused by the gravitational field of disk
and dark halo components, the initial radial and azimuthal
velocity dispersions are assigned to the disk component according to
the epicyclic theory with a Toomre's parameter $Q$. In the present study,
the Q parameters for stars ($Q_{\rm s}$) and gas ($Q_{\rm g}$)
are assumed to vary independently from one another meaning that
we can investigate how the initial stellar and gaseous kinematical properties of
dwarfs can influence the formation processes of GCs with multiple stellar populations.
The vertical velocity dispersion at a given radius is set to be half as large as the radial velocity dispersion at that point.

The interstellar medium (ISM) of the dwarf is modeled as a  thin gaseous disk
with the size of $R_{\rm g}=f_{\rm r}R_{\rm s}$, where $f_{\rm r}$ is a parameter
that determines the size ratio of gaseous to stellar  disks in a dwarf. We mainly
investigate the models with $f_{\rm r}=2$ in the present study.
The radial and vertical density profiles of the gas disk are
assumed to be proportional to $\exp (-R/R_{\rm g,0}) $ with scale
length $R_{\rm g,0} = 0.5R_{\rm g}$ and to ${\rm sech}^2 (Z/Z_{\rm g,0})$ with scale
length $Z_{\rm g,0} = 0.04R_{\rm s}$ , respectively.
Each gas particle is allocated an initial temperature ($T_{\rm g}$) and
the models with $T_{\rm g}=300$K, 1000K, and 10000K are investigated.
The radiative cooling processes
are properly included using the cooling curve by
Rosen \& Bregman (1995) for  $100 \le T < 10^4$K
and the MAPPING III code
for $T \ge 10^4$K
(Sutherland \& Dopita 1993).
The formation and evolution of dust and molecular (${\rm H_2}$) gas
that is properly modeled in our recent simulations for the  evolution of
gas-rich galaxies
(e.g., Yozin \& Bekki 2014; Cortese et al. 2016)
is not modeled in the present study.

We mainly show the results of the models in which all gas particles have initially 
the same metallicity ([Fe/H]$_0$). We discuss briefly how  initial metallicity gradients
in gas disks of dwarfs can influence the internal abundance spreads of FG stars of GCs later
in \S 4. Guided by a mass-metallicity relation ($Z \propto M^{1/4}$),
the initial metallicity of a dwarf is determined by the total stellar mass ($M_{\rm s}$) and 
the gas mass fraction ($f_{\rm g}$). For example, the standard model with 
$M_{\rm d}= 5 \times 10^8 {\it M}$ and $f_{\rm g}=0.8$ have [Fe/H]$_0=-1.45$. The values
of [Fe/H]$_0$ for $M_{\rm dm}=2 \times 10^9$, $6 \times 10^9$, and 
$6 \times 10^{10} M_{\odot}$ are $-1.83$, $-1.66$, and $-1.44$, respectively.
These initially metallicities are much less important, firstly because only radiative cooling
depends on [Fe/H] in the present study, and secondly because
the physics of clump formation (i.e., GC progenitors) is due to the 
dynamical instability of gas-rich dwarfs. 
We indeed confirmed
that the results do not depend on metallicities using models with
[Fe/H]=$-0.7$ and $-2.5$ as standard.

The initial total number of particles for dark matter halo, stellar disk, and gaseous disk
are $4 \times 10^5$, $4 \times 10^5$, and $2 \times 10^5$ in a dwarf disk galaxy.
The total number of particles can increase form this initial number ($N =10^6$) 
up to $N \sim 1.2 \times 10^6$ as new stellar particles  eject new gaseous particles
during their AGB phases. We need to investigate GC formation by using this $N \sim 10^6$,
because we have to finish running  $\sim 80$ models
for the limited amount of computational time allocated for this study.
The gravitational softening length ($\epsilon_{\rm g}$)
is assumed to be different between
different components (e.g., dark matter)
and determined from the initial mean particle separation for each component.
Therefore $\epsilon_{\rm g}$ depends both on the size and the mass of a dwarf, and the value
is later given for each model.

\subsection{Star formation and SN feedback}

A gas particle is converted
into a new star if the following condition is met:
\begin{equation}
\rho_{\rm g} \ge \rho_{\rm th},
\end{equation}
where $\rho_{\rm g}$ and $\rho_{\rm th}$ are
the local gas density around the gas particle
and  a threshold density for star formation.
The mass of the new stars is exactly the same as that of the original gas particle.
Although we investigate the models with $\rho_{\rm th}=1$, 10,
 100, and 1000 atoms cm$^{-3}$,
we show the results of the models with $\rho_{\rm th}=100$ atoms cm$^{-3}$, because GCs with compact
SG stellar systems can be clearly seen in these models. If $\rho_{\rm th}$ is much less than
100 atoms cm$^{-3}$, then the SG stellar systems in the present simulations become too diffuse
to be consistent with the observed GCs. 
On the other hand, if $\rho_{\rm th}$ is as large as 1000 atoms cm$^{-3}$, then star formation
is too strongly suppressed in gas disks leading to underdevelopment of FG stellar systems.
We thus need to adopt a reasonable $\rho_{\rm th}$ in the present galaxy-scale simulations,
because the simulations cannot resolve the real subparsec-scale star formation processes.
New stars formed from initial gas disks and from AGB ejecta
are referred to as FG and SG stars, respectively, for convenience.

A new star can become a SN and therefore
can eject gas and energy a certain time 
($t_{\rm sn}$) after its formation and the surrounding gas particles can receive the mass
and energy of the SN. 
The SN explosion can start $\sim 10^6$ yr after new star formation
and can continue until $\sim 3 \times 10^7$ yr after the star formation.
These values are reasonable, given the lifetime of the least and most massive
progenitors of SNII ($8 M_{\odot}$ and $100 M_{\odot}$).
Thornton et al. (1998) investigated
the energy conversion processes of SNe in  detail
and found that about $8.5 \times 10^{49}$ ergs among the total energy of a SN
($\sim 10^{51}$ erg) can be in the form of kinetic energy.
Following these results,
we consider that (i)
the energy of each SN is assumed to 
be used for the increase
in random motion (`kinematic feedback') in the present study
and (ii)
the ejected gas with an ejection velocity ($V_{\rm ej}$) of 920 km s$^{-1}$
(corresponding to less than 10\% of the initial SN energy)  can  be  mixed with
the surrounding gas particles soon after SN explosion. 
We also investigate `stronger feedback models' with $V_{\rm ej}=2916$  km s$^{-1}$ to understand
how the present results depend on the modeling of SN feedback effects.
We consider only SNII (not SNIa) in the 
present study, mainly because we investigate only 560 Myr evolution of dwarf galaxies. 
The canonical Salpeter IMF (the slope of $\alpha_{\rm IMF}=2.35$)
with the lower and upper cut-off masses being $0.1M_{\odot}$ and
$50 M_{\odot}$, respectively,  is adopted and the number of SNII per unit mass
is calculated for the adopted IMF.

   \begin{figure*}
   \centering
   \includegraphics[width=14cm]{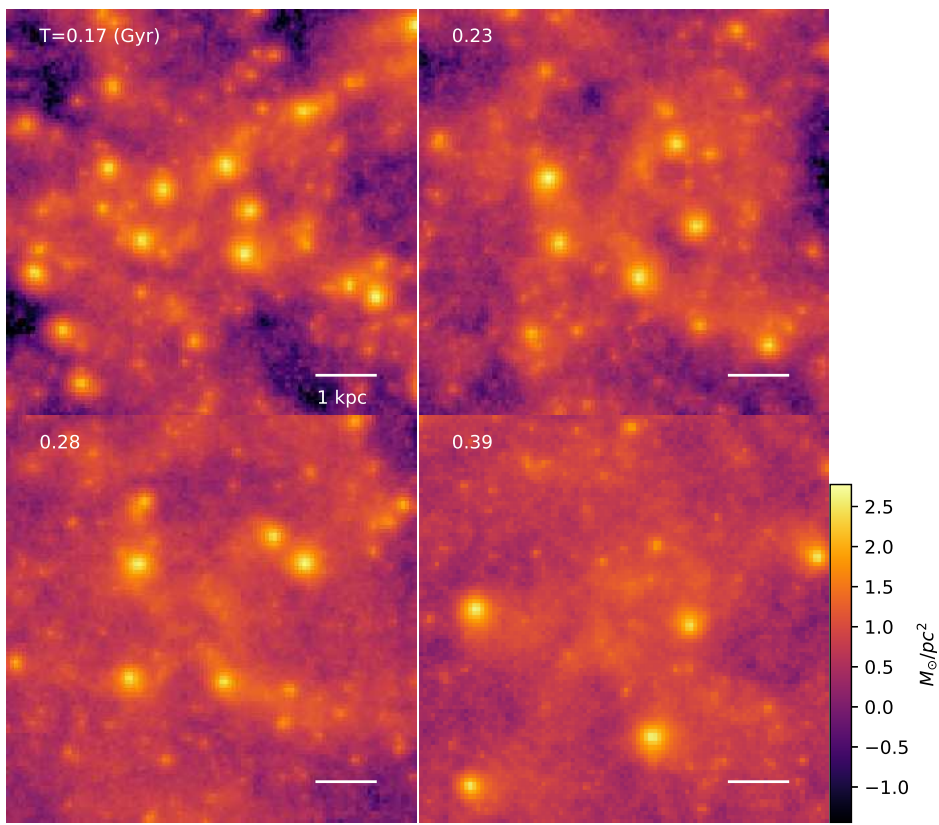}
      \caption{
Time evolution of stellar 
 distributions 
projected onto the $x$-$y$ (left) plane
in the standard model. Clearly, several massive
high-density  clumps of FG stars, where SG stars
are forming, can be formed within 0.3 Gyr. 
              }
         \label{FigVibStab}
   \end{figure*}

\subsection{A new model for the evolution of AGB ejecta}

The mass of a particle that is used for modeling (i) ejection of gas
from AGB stars and (ii) the subsequent star formation from the gas
is much smaller than the masses of old stars and dark matter halos.
Furthermore, feedback effects of AGB wind ($\sim [10-20]$ km s$^{-1}$)
need to be properly modeled within a scale of less than 100 pc. 
Accordingly,
we adopt an original numerical method
for gas ejection and feedback effects of AGB stars
in order to investigate this important secondary star formation from AGB ejecta
with  sufficient mass and scale resolutions.
 In the present simulations,
we try to resolve pc-scale dynamical evolution of AGB ejecta within newly
formed FG stellar systems by (i) ejecting   `AGB' gas particles with masses
much smaller than the initial gas particle masses
within local FG systems and  (ii) adopting very short time-step width
and significantly smaller gravitational softening length for the AGB gas particles.
This `AGB gas ejection method' has a number of advantages in simulating
GC formation from massive gas clumps,  which are  described later in this paper.
The details of the method are given as follows.

\subsubsection{Ejection of new particles}

Each new star is assumed to
eject $n_{\rm agb}$ new gaseous (SPH) 
particles when it enters into  its  AGB phase,
and these gaseous particles correspond to AGB ejecta and are referred to
as `AGB particles' for convenience.
The mass of AGB ejecta ($m_{\rm agb}$)
is much smaller than the original gas particle mass,
and many particles ($\sim 1000$)
from many new stars can be generated in a forming FG stellar
system. Therefore, the mass-resolution in the FG system can be much better
than that for the host dwarf galaxy.
The value of $m_{\rm agb}$ is determined according to the adopted
IMF and the mass range of AGB stars. For example,
$m_{\rm agb} \sim 0.05 m_{\rm g}$ for a standard IMF, where
$m_{\rm g}$ is the mass of an original gas particle.
Thus, the mass resolution of a forming GC in the present study
can be as good as $[10^2-10^3] {\rm M}_{\odot}$: the resolution depends on
the initial total mass of the dwarf and the initial particle number of gas.

Although it is ideal for the present simulations
to model continuous ejection of gas during AGB phases
(over many time steps)  for different AGB stars with different masses,
it is extremely numerically costly and indeed impractical to simulate such continuous AGB
gas ejection because of a huge number of particles required for such simulations.
Therefore, we assume that each new star can eject one new particle ($n_{\rm agb}=1$) with
a mass $m_{\rm agb}$ 
only one at a time ($t_{\rm agb}$) 
when it becomes
an AGB star.  
These AGB particles are assumed to interact gravitationally and hydrodynamically with
neighboring particles.
By using 
the nucleosynthesis yields of AGB stars
from van den Hoek \& Groenewegen (1997, VG97), we calculate $m_{\rm agb}$ for the adopted Salpeter IMF
in each model. 
Since we consider that AGB stars with their initial masses ranging from
$5 M_{\odot}$  to $8 M_{\odot}$ can eject gas that can be used for the formation
of SG stars,
we assume that the lifetime of a star with $m=8M_{\odot}$ can correspond
to $t_{\rm agb}$.  We do not model the continuous ejection of gas from AGB stars
with different masses (i.e., do not allocate different $t_{\rm agb}$ for different particles)
owing to the above-mentioned numerical cost (i.e., a huge number of particles necessary to
model this).

This adopted AGB gas ejection method has the following additional
advantages in simulating
GC formation within dwarfs. 
AGB gas particles can interact with other new AGB gas particles
within FG stellar systems so that star formation from pure AGB
ejecta and gas dynamics of AGB ejecta  within the FG systems can be self-consistently
investigated. In standard galaxy-scale chemodynamical simulations,
all AGB ejecta are assumed to mix with interstellar medium without generating
new AGB particles. Therefore,  star formation from pure AGB ejecta cannot be properly modeled.
Thus,
if there is no AGB gas particle in numerical simulations,
then we cannot investigate
crucial physical processes of GC formation
(i.e., star formation from pure AGB ejecta)  using the simulations.

\subsubsection{AGB feedback effects}

New particles are ejected
from AGB stars with ejection velocities ($v_{\rm ej, agb}$) and gaseous temperatures
($T_{\rm g, agb}$) meaning that they can influence local dynamics of gas around AGB stars.
As shown by previous numerical simulations (D08; B11),  
AGB ejecta with $v_{\rm ej, agb} \sim 10$ km s$^{-1}$ cannot  
be kept in less massive  FG stellar systems. It is thus possible that AGB ejecta can also significantly
influence the formation processes of SG stars in the present simulations.  We investigate
this `AGB feedback effect' on the formation and evolution of FG and SG stellar systems
mainly for $v_{\rm ej, agb}=20$ km s$^{-1}$ and $T_{\rm g, agb}=1000$K.
We also investigate the model without AGB feedback effects ($v_{\rm ej, agb}=0$ km s$^{-1}$)
in order to understand more clearly whether stellar winds of AGB stars can suppress star formation
at GC formation.

\subsubsection{Much smaller individual time step width}

We consider that 
the maximum time step width ($\delta t_{\rm max}$) should be different between
AGB particles and other particles in a simulation. 
For dark matter particles, stellar particles, and gaseous particles (other than AGB ejecta),
$\delta t_{\rm max}$ is set to be $1.4 \times 10^6$ yr 
and the time step width at each time step is determined for each particle according to
the physical conditions  of the particle (e.g., Courant condition) for all
models in the present study. This $\delta t_{\rm max}$ is not short enough to properly investigate
the formation of SG stars from AGB ejecta in the FG stellar systems, because the local dynamical time
scale of the FG systems is significantly shorter than $\sim 10^6$ yr. We therefore consider that
AGB ejecta particles can have  $\delta t_{\rm max}$ ($\delta t_{\rm max, agb}$) significantly shorter
than $1.4 \times 10^6$ yr. We mainly show the results with 
$\delta t_{\rm max, agb}=8.8 \times 10^4$ yr, because the models with 
$\delta t_{\rm max, agb} \le 8.8 \times 10^4$ yr show rather similar results on the formation
of SG stellar systems.

It is confirmed that if $\delta t_{\rm max, agb}=1.4 \times 10^6$ yr
(corresponding to the time step width for original gas and star particles),
then compact SG stellar systems cannot be formed.
This demonstrates that a much smaller individual time step width
is required for simulating the formation of SG stars from AGB ejecta.
The adopted time-stepping method for AGB ejecta within
forming star clusters is quite different from those adopted in previous
galaxy-scale simulations including ours (e.g., Bekki 2013).
It should be stressed here that this kind of time-stepping method
is necessary to discuss the evolution of ejecta  from dying stars and SNe
within existing star clusters.
Owing to the much smaller time step width
(and many AGB particles within a forming FG system), radiative cooling 
based on the cooling curve by Rosen \& Bregman (1995) for $T<10^4$ K 
can be properly included for the SG stars within the FG system.
Thus, this adopted very small time-step width for AGB particles enables the
present simulations to properly investigate (i) whether AGB ejecta can 
escape from forming GCs and (ii) whether AGB ejecta can be converted
into new stars in galaxy-scale simulations for the first time .

\subsubsection{Gravitational softening length}

We have to adopt both short $\delta t_{\rm max}$ and small $\epsilon_{\rm g}$ for AGB ejecta particles
to properly investigate the formation of SG stellar systems in forming GCs, because the SG systems
should be rather compact ($\sim 10$pc). We therefore consider that $\epsilon_{\rm g}$ for new stars
formed from gas (either from initial disk gas or from AGB ejecta) should be significantly smaller
than that for old stars. We assume that $\epsilon_{\rm g}$ for new stars ($\epsilon_{\rm g, ns}$)
is 10\% of $\epsilon_{\rm g}$ for old stars ($\epsilon_{\rm g,os}$). 
The value of $\epsilon_{\rm g, ns}$ is typically $\sim 2$pc for dwarfs with 
$M_{\rm s} \sim 10^9 M_{\odot}$.
Thanks to this 
small $\epsilon_{\rm g,ns}$ combined with short $\delta t_{\rm max, agb}$, we can better investigate
the formation processes of GCs with FG and SG stars in the present study.

The gravitational softening lengths for massive dark matter ($\epsilon_{\rm g, dm}$),
old disk stars ($\epsilon_{\rm g,os}$),  ISM 
($\epsilon_{\rm g, g}=\epsilon_{\rm g, os}$), new stars ($\epsilon_{\rm g,ns}$),
and AGB ejecta ($\epsilon_{\rm g,agb}=\epsilon_{\rm g, ns}$) are quite different
in the present study.
When two different components interact gravitationally,
the mean softening length for the two components
is applied for the gravitational calculation.
For example, $\epsilon_{\rm g}$ for
the gravitational interaction between old stars and SG stars (and AGB ejecta) is as follows:
\begin{equation}
\epsilon_{\rm g} = \frac{ \epsilon_{\rm g,os}+{\epsilon}_{\rm g, ns} }{2}.
\end{equation}
Although the softening length
of dark matter particles
is relatively large (and therefore the spatial resolution is poorer),
the adopted multiple softening lengths (much smaller softening length
for SG stars from AGB ejecta)
can avoid unrealistic dynamical heating of 
SG stars by old stars and dark matter halos
in dwarfs, and guarantee that linear and angular momentum can be conserved
for a simulation with different softening lengths.

   \begin{figure*}
   \centering
   \includegraphics{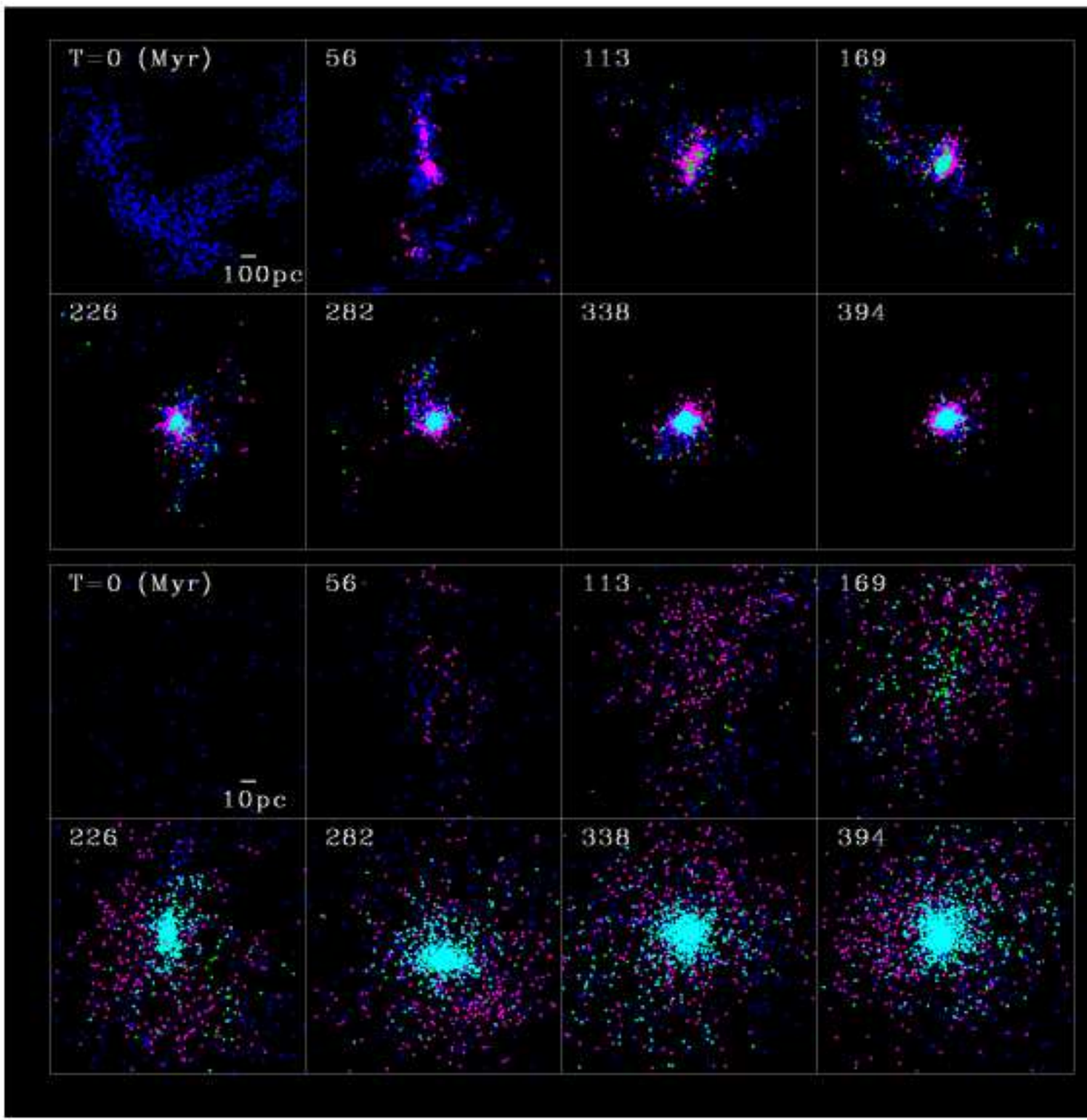}
   \caption{
The time evolution of mass distributions projected onto the $x$-$y$ plane
for ISM (blue),
FG stars (magenta),  AGB ejecta (green), and SG stars (cyan) of GC1
in the standard model (M1).
The upper and lower  eight panels are for larger and smaller scales of view, respectively.
The time $T$ in the upper left corner of each panel shows the time (in units of Myr)
that has elapsed since
the simulation stars. The thick bar indicates 100pc and 10pc for upper and lower eight panels,
respectively. The mass-center of GC1 is the center of each frame in this figure.
}
              \label{FigGam}%
    \end{figure*}

\subsubsection{Star formation of second generation stars}

We consider that the threshold SF density can be different between SF
in gas clouds (i.e., in `normal' situations) and in dense stellar systems (i.e., in FG stellar systems),
because high-velocity interaction between stars and forming molecular gas clouds in FG stellar systems
could prevent the gas clouds from collapsing gravitationally. 
Therefore $\rho_{\rm th}$ for SG formation ($\rho_{\rm th, SG}$ for convenience) can be different from
$\rho_{\rm th}$ for FG star formation. We mainly investigate the models in
which $\rho_{\rm th,SG}=\rho_{\rm th}$ in the present study and discuss briefly how
$\rho_{\rm th, SG}$ can influence the formation of compact SG stellar systems in forming GCs.
The new stars formed from AGB ejecta (i.e., SG stars) are assumed to eject no gas during their AGB phase
meaning that the total number of particles cannot dramatically increase owing to the formation of
third/fourth generations
of stars.  This assumption is reasonable, because the total mass of new stars formed from SG AGB stars
is much smaller than the total mass of FG and SG stars.

\subsection{Identification of globular cluster candidates}

It is an important task for this study to precisely identify GC candidates with both FG and SG
stars for each model. The method to identify GC candidates is as follows.
First, we investigate the total masses of FG stars ($M_{\rm FG}$), 
ISM ($M_{\rm ISM}$), AGB ejecta ($M_{\rm AGB}$), and SG stars ($M_{\rm SG}$) 
within $r_{\rm GC}$ from the location
of  each new stellar particle at the final time step ($T=0.56$ Gyr) in each model.
The total mass of a GC for GC identification  is therefore denoted as:
\begin{equation}
M_{\rm GC}=M_{\rm FG}+M_{\rm ISM}+M_{\rm SG}+M_{\rm AGB}.
\end{equation}
If $M_{\rm GC}$ exceeds a threshold mass ($M_{\rm th, GC}$) beyond which GCs with
the present typical mass  of $2 \times 10^5 M_{\odot}$ are considered to be able to form,
then the new stellar particle is regarded as being in a massive GC-like system (or clump).
In this way, we try to identify 
all massive GC-like systems in each model.
We adopt $M_{\rm th, GC}=2 \times 10^6 M_{\odot}$ and $r_{\rm gc}=57$ pc
as reasonable values to clearly
identify originally massive GC-like systems.
Since some GC-like systems 
with  $M_{\rm GC} \ge M_{\rm th, GC}$ can be dominated by ISM and AGB ejecta,
we have to select real GC candidates from the selected massive systems.

Subsequently, we investigate $M_{\rm SG}$
in each of the selected  massive GC-like systems,
and regard the system as a {\it GC candidate}
if $M_{\rm SG} \ge 2 \times 10^5 M_{\odot}$.
The stellar  systems with $M_{\rm SG} < 2\times 10^5 M_{\odot}$ 
are likely to become low-mass clusters  without
significant SG populations after most of the FG stars are lost after SG formation.
These systems are found to have very diffuse FG stellar systems for most cases,
and some of them have a larger amount of gas (i.e., identified as compact systems owing to
high gas densities).
Although the majority of GC candidates identified as above 
can have compact SG stellar systems,
some of them can have diffuse ones, which can only be confirmed by investigating
morphological properties and radial density profiles
(i.e., not automatically).  These GC candidates might  not be regarded as `genuine GC',
but the properties of these GCs are used for some statistical discussion on GC properties.

   \begin{figure}
   \centering
   \includegraphics[width=8cm]{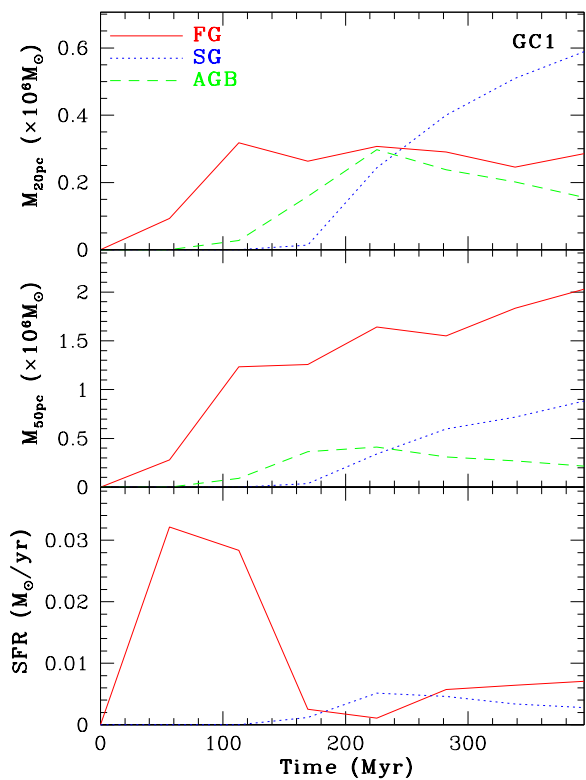}
      \caption{
 The time evolution of total mass within 20pc (top) and 50pc (middle) for
FG (red solid),  SG (blue dotted), and AGB (green dashed) stars for GC1 in the standard model.
The bottom panel shows the mean SF rate at selected time steps for FG (red solid) and SG (blue dotted)
              }
         \label{FigVibStab}
   \end{figure}

\subsection{Parameter study}

We mainly describe the results of the `standard model' with 
$M_{\rm dm}=2\times 10^{10} M_{\odot}$ within $5a_{\rm dm}$ (i.e., $f_{\rm b}=0.11$),
$M_{\rm s}=5\times 10^{8} M_{\odot}$,
$M_{\rm g}=2\times 10^{9} M_{\odot}$ (i.e., $f_{\rm g}=0.8$),
$a_{\rm dm}=2.8$ kpc,
$R_{\rm s}=2.3$ kpc,
$R_{\rm g}=4.6$ kpc,
$Q_{\rm s}=1.5$,
$Q_{\rm g}=0.5$,
$T_{\rm g}=300$K,
$\rho_{\rm th}=\rho_{\rm th, SG}=100$ atoms cm$^{-3}$,
[Fe/H]$_0=-1.45$,
$\epsilon_{\rm g,dm}=193$pc,
$\epsilon_{\rm g,os}=21$pc,
and $\epsilon_{\rm g,ns}=2$pc,
because this model 
more clearly shows essential ingredients
of the sequential formation processes of FG and SG stars in forming GCs.
A summary of these model parameters is given in Table 1.
 It should be noted here that $f_{\rm b}$ is high in each model 
because dark matter halo is truncated at
$r = 5a_{\rm dm}$. If the halo is extended to the virial radius,
then $f_{\rm b}$ should be significantly higher than the adopted value.
Although we have investigated many models ($\sim 100$) in order to find models in
which {\it genuine} GCs with both FG and SG stars are formed,
we describe the results only for  key representative models in the present study.
The parameter values for these 29 models are summarized in Table 2.

Dwarf galaxies are observed to have higher gas fractions in the local
universe.
For example,
the following correlation between
gas (neutral hydrogen) mass ($M_{\rm g}$) and stellar mass ($M_{\ast}$)
in galaxies with different masses and types at low redshifts 
is derived by
Papastergis et al. 2012:
\begin{equation}
\log(M_{\rm g}/M_{\ast})  = -0.43 \log(M_{\ast}/M_{\odot}) + 3.75.
\end{equation}
Using this relation, a reasonable gas mass fraction for the 
dwarf with $M_{\rm s}=5 \times 10^8 M_{\odot}$ adopted in the standard model
is estimated to be 1.02 ($f_{\rm g}\sim 0.5$).  The adopted $f_{\rm g}$
of 0.8 in the standard model
is significantly higher than the above value,
though the observed relation shows a large scatter for a given stellar mass.
Given that high-$z$ galaxies are observed to have higher gas fractions
(e.g., Narayanan et al. 2012 for detailed discussion),
the adopted gas fraction should
be reasonable for high-$z$ dwarf galaxies.

Figure 2 shows the initial rotation curve profile of the dwarf disk galaxy
and the gas distribution projected onto the $x$-$y$ plane        
at $T=56$ Myr
in the standard model.
The model shows a slowly rising rotation curve until $R=3$ kpc
owing to the adopted cored dark matter halo,
and the lower dark matter density in the inner region means
a higher degree of self-gravitation in the baryonic components,
which can possibly
play a vital role in the evolution of the gas disk.
The shapes of rotation curves do not evolve significantly within 0.56 Gyr
owing to a lack of merging and gas infall.
The dwarf disk is initially so gas-rich ($f_{\rm g}=0.8$)
that  many filamentary    
or clumpy structures can develope 
as a result of local gravitational  instability.
Formation of FG and SG stellar systems from these gaseous structures
is the most important issue in the present numerical simulations.

In the present study,
$M_{\rm dm}$, $f_{\rm g}$, and $f_{\rm b}$
are considered to be the key parameters that mainly determine whether GCs with compact 
SG stellar systems can be formed. 
We therefore describe the results of  the models with 
$2 \times 10^9 \le M_{\rm dm}/ M_{\odot} \le 6 \times 10^{10}$,
$0.02 \le f_{\rm b} \le 0.11$,
and $0.2 \le f_{\rm g}  \le 0.95$.
We also show the results of the `LSB' (low surface-brightness) models in which
the initial mean surface mass densities of  the  stellar disks are $2.5^2$ times lower
in comparison with  the standard model that is regarded as a HSB galaxy.
The `higher surface density model' has 
a stellar disk two times smaller than that of the standard model.

\subsection{Limitations of the model}

Gas particles ejected from AGB stars can be converted into new stars without mixing
with pristine ISM around forming GCs in the present simulations.  Although this method
to convert AGB ejecta into new stars allows us to investigate how SG stars can be formed
from AGB ejecta,  we are unable to discuss how the mixing of AGB ejecta and ISM can
determine the chemical abundances of SG stars. Also, ISM that is later accreted onto
the inner regions of forming FG stellar systems can be identified as `FG' stars
in the present model, though such ISM can be mixed with AGB ejecta and then converted
into SG stars. This is one of the major limitations of the present study, and accordingly
we need to consider this in interpreting the simulation results.

AGB stars with different masses can eject different amounts of gas with different
chemical compositions, which can determine the nature of (anti-) correlations between
different chemical abundances within GCs. This means that a large number of gas particles
would need to be used for just one stellar particle consisting of stars with
different masses for long-term chemical enrichment of intra-cluster medium by
AGB stars. In the present study, we investigate the models with $n_{\rm AGB}=1$
in order to avoid the expected huger number of gas particles. 
This is one reason why the present simulations do not allow us to discuss the details
of chemical abundances of SG stars. We will need to more properly incorporate
the AGB ejecta with different chemical compositions in our future simulations.

Furthermore, the resolution of each simulation is only 2pc at most and gravitational
softening lengths are  applied.  This means that we cannot discuss the long-term
evolution of compact stellar systems through two-body dynamical relaxation processes,
which can be properly investigated by the NBODY 6 code (e.g., Hurley \& Shara 2012).
Therefore, the simulated GCs with number densities of $10^4 - 10^5$ stars pc$^{-3}$  at most
(in particular
FG stellar systems) look more diffuse than the real GC.
In order to obtain a fully self-consistent model of GC formation from their birth
to destruction, we need to develop a code with which we can investigate both
galaxy-scale hydrodynamics of cold gas and dynamical evolution of forming clusters.
Clearly, this is beyond the scope of this paper.

   \begin{figure}
   \centering
   \includegraphics[width=8cm]{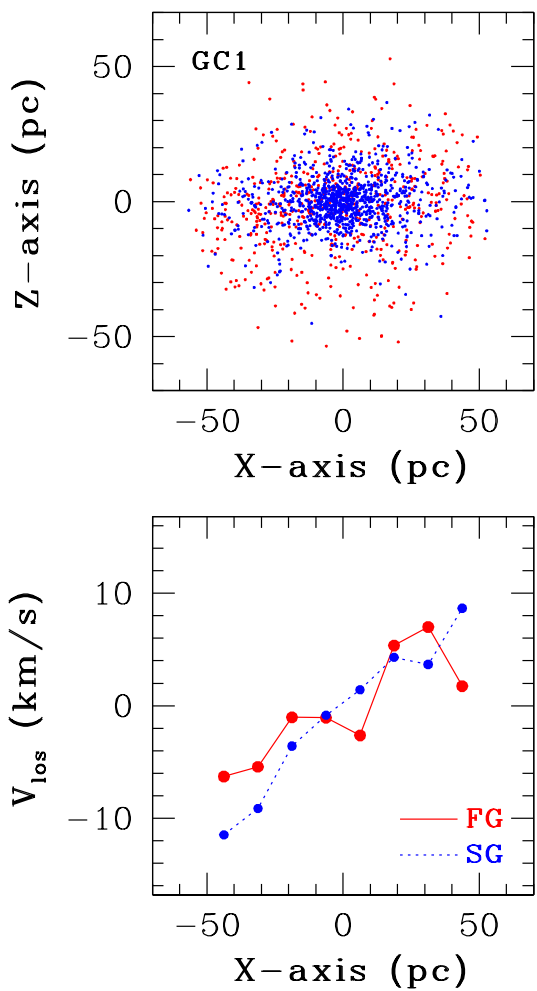}
      \caption{
The distributions of FG (red) and SG (blue) stars projected onto the $x$-$z$ plane (upper)
and the line-of-sight velocity profile ($V_{\rm los}$) for FG (solid) and SG (dotted) stars
(lower) for GC1 in the standard model.
              }
         \label{FigVibStab}
   \end{figure}

\section{Results}

\subsection{Standard model}

Figure 3 shows the time evolution
of the spatial distribution of new stars 
in the standard model. Clearly, the dwarf disk galaxy can develop
several massive high-density stellar clumps, where FG stellar systems
(GC progenitors) can form.
In the standard model, seven GC candidates (GC1$-7$) with $M_{\rm GC} \ge 2 \times  10^6  M_{\odot}$
 are identified in the stellar disk of an initially gas-rich dwarf at the final
time step ($T=560$ Myr). 
Among these, GC1,  GC2, GC4, and GC6 can be regarded as genuine GCs,
because they have  $M_{\rm SG} \ge 2 \times 10^5  M_{\odot}$ and a compact SG stellar system.
Accordingly, the 3D locations for only  these four genuine GCs are shown in this figure.
The same large stellar system can be  identified
twice as a GC candidate, and the center of mass  
can be  very similar between two GC candidates. GC2 and GC3 have very similar 
center of mass  positions,
therefore GC3 was removed from the list of genuine  GCs in this model (to avoid double GC counts).
None of these GCs with FG and SG stars contain dark matter particles within
them ($R<r_{\rm gc}=$ 57pc).

   \begin{figure}
   \centering
   \includegraphics[width=8cm]{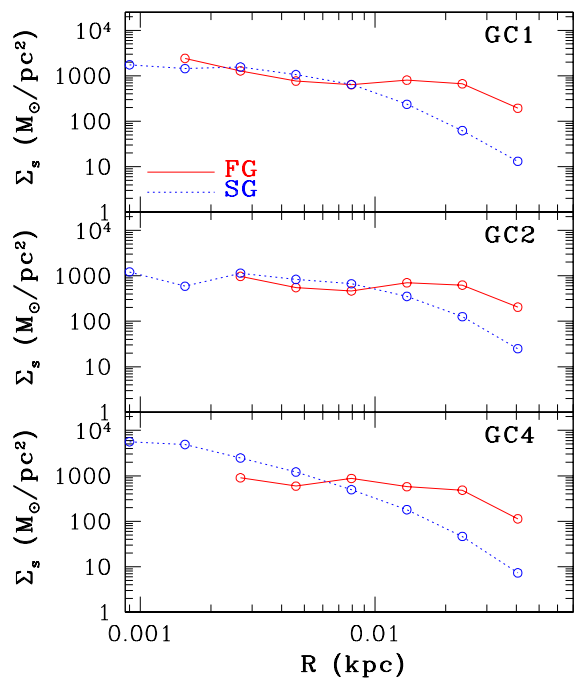}
      \caption{
The projected radial density profiles ($\Sigma_{\rm s}$) of FG (red solid) and SG (blue dotted)
stars for GC1 (top), GC2 (middle),  and GC4 (bottom) in the standard model.
              }
         \label{FigVibStab}
   \end{figure}

\subsubsection{Two-stage GC formation}

Figure 4 shows essential ingredients of GC formation processes in a gas-rich dwarf galaxy for GC1
with $M_{\rm FG}=4.7 \times 10^6  M_{\odot}$ and 
$M_{\rm SG}=9.6 \times 10^5  M_{\odot}$  at $T=394$ Myr
in the standard model.  The ISM from which  GC1 can form
comes originally from different regions widely spread in
the initial gas-rich disk meaning that GC1 can finally become quite massive. 
During the dynamical evolution of the gas-rich disk  in the dwarf,  new stars can form from
the high-density regions of 100pc-scale filamentary structures
that are developed through local gravitational instability 
(owing to lower $Q_{\rm g}$). These stars can become
two massive stellar clusters in the filaments  and field stars ($T=56$ Myr). 
The two clusters correspond
to the progenitor of the FG stellar system for GC1. When the two clusters are about to 
merge ($T=113$ Myr),  some of the FG stars
enter into their AGB phases and consequently start to eject gas. In the late phase of
cluster merging ($T=169$ Myr),  the AGB ejecta can be trapped by the cluster potential and start
being converted
into new stars (i.e., SG stars).

After the formation of the FG stellar system by merging of clusters,
ISM and  AGB ejecta can continue to be accumulated in the central region of the diffuse FG stellar 
system and converted into new stars with a high SF efficiency ($T=226$ Myr).
It should be noted that these gases are located in filamentary or tail-like structures before their
accretion onto the FG stellar system.
A compact and elongated SG stellar system can be developed ($T=282$ Myr), and the  accretion process
of AGB ejecta and ISM can continue after the formation of this nested stellar system.
The infant GC with FG and SG stars can finally have a rounder shape ($T=394$ Myr) owing to
dynamical relaxation processes. For this massive GC, ISM that is not pushed out by SNII
can be later accreted onto the diffuse FG stellar system to be mixed with AGB ejecta and 
converted into new stars. This implies that accretion of ISM onto already
existing FG stellar systems
is a key process for dilution of AGB ejecta in the formation of SG stars. 

Figure 5 shows that the major formation epoch of FG stars is $\sim 150$ Myr earlier than that of
SG stars in this model. However, the formation of FG stars 
can continue even after the major epoch of SG stars ($T\sim 230$ Myr) for this massive
GC. Some of these FG stars forming later than $T=230$ Myr cannot be regarded
as genuine FG stars, because they form after mixing of interstellar gas with AGB ejecta.
Although the total mass of SG stars within the central 20pc of GC1 is smaller than that
of FG stars in the early GC evolution, it can finally become significantly
larger than that of FG stars at $T\sim 300$ Myr.
This result means that a nested GC structure can grow on a timescale of $\sim 200$ Myr.
The total mass of FG stars within 50pc is larger than
that of SG stars by a factor of approximately two, which means that most FG stars need to be lost for this GC1 to become
similar to the present GCs dominated by SG stars.

AGB  ejecta and SG stars of GC1 at $T=394$ Myr do not necessarily originate from FG AGB stars
of GC1 in this model: About 39\% of AGB ejecta in GC1 are from AGB stars that are {\it not}
within GC1. This means that although some AGB stars formed from ISM in the dwarf disk
eject gas that can finally contribute to the formation of 
SG stars in GC1, they cannot finally become member stars of GC1. 
Recent theoretical models of GC formation with multiple stellar populations (e.g., D08; B11)
have adopted an assumption that all SG stars are formed from AGB ejecta of FG stars.
The present result therefore suggests that (i) such an assumption in these recent models
is over-simplified and less realistic and (ii) gas from field AGB stars can also be important
for the formation of SG stars.

The GC formation processes described above are essentially similar to those investigated in
previous studies (e.g., D08; B11) in that most SG stars can form
from AGB ejecta in diffuse FG stellar systems that form about 100-200 Myr before the major epoch
of SG formation.  It is confirmed that
this two-stage GC formation is not only for GC1 but also for most of the simulated GCs
in the present study. FG stars are initially in filamentary or clumpy structures including sub-clusters,
and merging of the clusters is essential for the formation of FG stellar system.
Although the majority 
of SG stars can form after cluster merging for  GC1 in
the standard model and most of other models,  SG formation can occur in two different clusters
before the clusters merge with each other in some models. Below, we discuss this later merging
of GCs with SG stars in the context of binary GC formation.

\subsubsection{Kinematics and structures of GC candidates}

Figure 6 shows the final mass distributions of FG and SG stars and line-of-sight rotation
curve profiles ($V_{\rm los}$) for GC1.  The SG stellar system appears to be flattened and
has a larger amplitude of rotation ($\sim 10$ km s$^{-1}$) than the FG system. 
The $V_{\rm los}$ profile for FG stars appears to change more violently at some radii (e.g., at
10pc in the $x$-axis), because the FG system contains new stars that are captured later by
GC1  and therefore have stream motions within the GC (i.e., not necessarily bound by GC1). 
The estimated  $V/\sigma$, where $V$ and $\sigma$ are the maximum $V_{\rm los}$ and velocity dispersion,
respectively, is 0.34 for FG stars and 0.70 for SG stars for GC1.
This larger $V/\sigma$ in SG stars can be seen in other GCs in the standard and other models. 
This result is consistent with our previous works (Bekki 2009; B11), which have already shown
that SG stellar systems are  more strongly dynamically supported  by rotation at their formation.
It should be stressed that the FG systems of GC1 and other
massive GCs  can have rotational kinematics owing
to merging of sub-clusters at the early formation phases of the systems in the present study.
This result implies that the origin of rotation observed in some Galactic GCs 
(e.g., Meylan \& Mayor 1986; Anderson \& King 2003; Pancino et al. 2007) 
can be closely related to early formation
processes of FG stellar systems through cluster merging.

Figure 7 shows that the projected radial density profiles are steeper in SG stars than in FG stars
for GC1, GC2, and GC4.  The simulated structures of FG stars in the present study
is  much less compact than those modeled in previous studies (e.g., D08 and  B11),
which implies that the initial conditions for FG stellar systems in these previous studies
may not be  realistic.
The central density at $R\sim 1$pc can be larger than $10^3  M_{\odot}$ pc$^{-2}$ 
but less than $10^4  M_{\odot}$ pc$^{-2}$ in SG stars  for these three GCs. The inner density
profiles of SG stars have flat cores in the three. These central structures are due largely
to the adopted gravitational softening length ($\epsilon_{\rm g} \sim 2$ pc) for SG stars.
The softening length is too large for this study to properly investigate the dynamical structures
for the inner regions ($R<1$pc) of the simulated GCs. The present code does not allow us to investigate
the long-term dynamical evolution of GCs driven by two-body relaxation processes within GCs
owing to the introduction of  a gravitational softening length. Future numerical simulations
using a proper code (e.g., NBODY6) will help us to better understand the final structures
of GCs with FG and SG stars after their long-term ($1-10$ Gyr) dynamical evolution.

The simulated nested structures of GCs with different radial profiles 
between FG and SG stars imply that GC stars with different ages can have
different radial density profiles. It is, however, observationally difficult
to separately investigate the radial density profiles of stellar populations
with different ages in old GCs.
Recently, Li et al. (2016) discovered two stellar populations
with different ages of a few hundred million years 
in three intermediate-age GCs within the LMC and thus confirmed
that there was secondary star formation possibly from accreted gas onto the GCs.
The above simulation results suggest that
if the radial profiles of the two populations in each of these
GCs and other LMC GCs with age spreads among the stars
(e.g., Girardi et al. 2011; Goudfrooij et al. 2014; Milone et al. 2015) 
can be derived and then compared  with the corresponding simulations,
then the origin of these GCs will be better understood.

   \begin{figure}
   \centering
   \includegraphics[width=8cm]{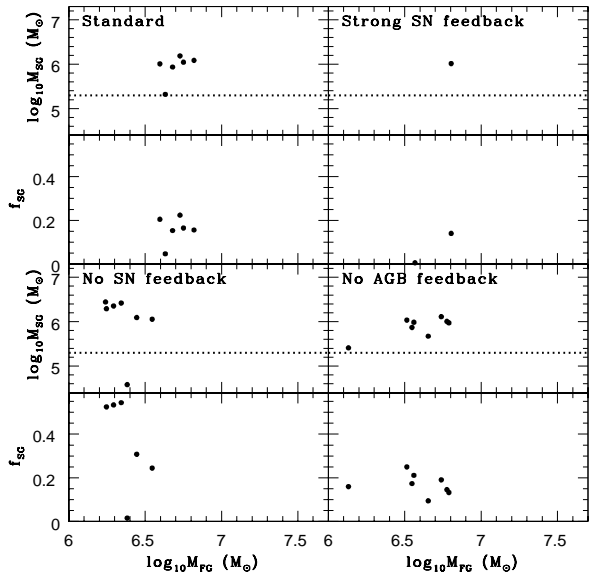}
      \caption{
The dependencies of $M_{\rm SG}$ and $f_{\rm SG}$ on $M_{\rm FG}$ for
GC candidates in four models:
the standard model M1 (upper left),  M2 with strong SG feedback effects (upper right),
M3  with no SN feedback effects (lower left), and M4 with no AGB feedback effects (lower right).
              }
         \label{FigVibStab}
   \end{figure}

   \begin{figure}
   \centering
   \includegraphics[width=8cm]{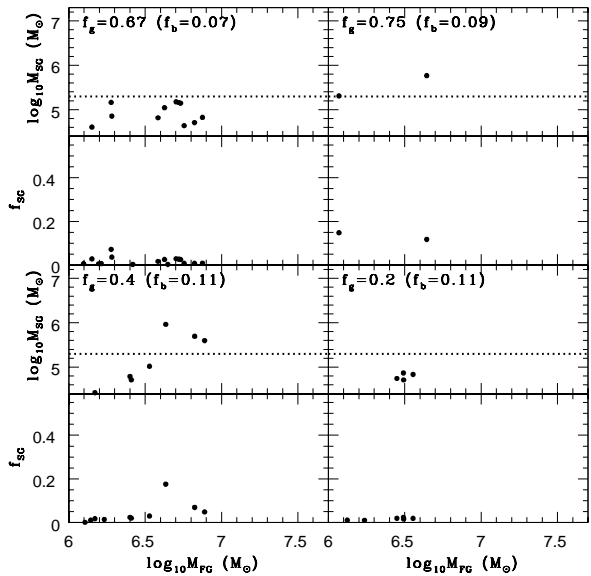}
      \caption{
As in Figure 8 but for
GC candidates in four models with different $f_{\rm g}$ (and $f_{\rm b}$):
M5 (upper left),  M6 (upper right),
M7 (lower left), and M8 (lower right).
Other model parameters for these four models are the same as those in the standard model.
              }
         \label{FigVibStab}
   \end{figure}

   \begin{figure}
   \centering
   \includegraphics[width=8cm]{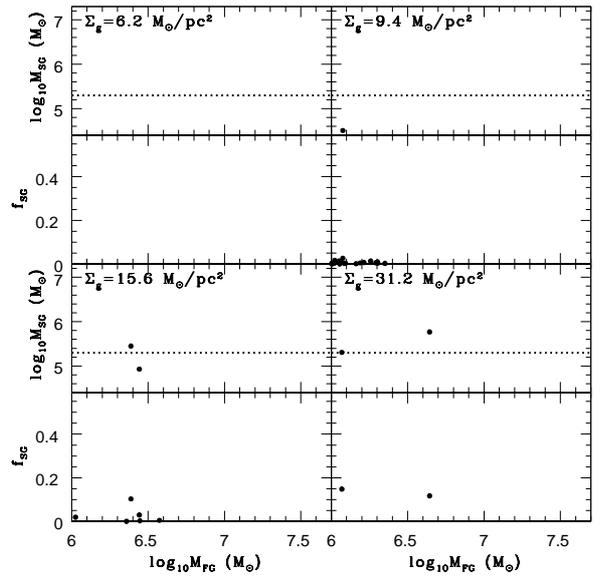}
      \caption{
As in Figure 8 but for
GC candidates in four models with different $\Sigma_{\rm g}$ (i.e., initial mean gas
surface density):
M9 (upper left),  M10 (upper right),
M11 (lower left), and M12 (lower right).
The initial stellar mass is the same ($10^8 {\it M}_{\odot}$) between the four models
whereas the initial gas mass is different for different $\Sigma_{\rm g}$.
Other model parameters for these four models are the same as those in the standard model.
              }
         \label{FigVibStab}
   \end{figure}

   \begin{figure}
   \centering
   \includegraphics[width=8cm]{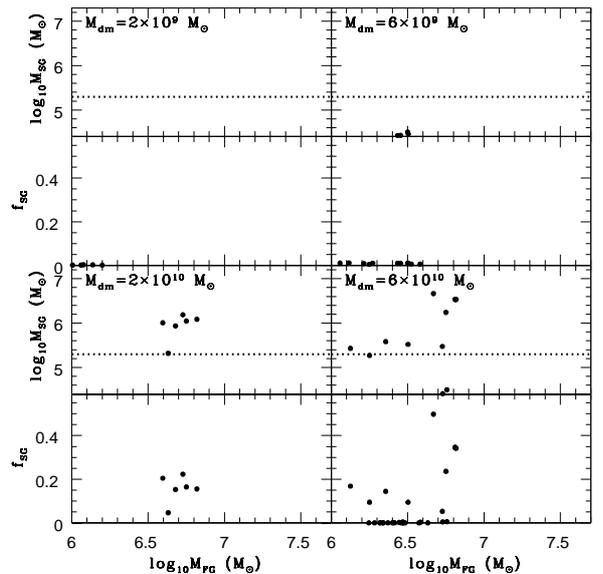}
      \caption{
As in Figure 8 but for
GC candidates in four models with  different $M_{\rm dm}$:
M16 (upper left),  M17 (upper right),
M1 (lower left), and M18 (lower right).
$f_{\rm g}$ and $f_{\rm b}$ are the same between these four models.
              }
         \label{FigVibStab}
   \end{figure}

   \begin{figure}
   \centering
   \includegraphics[width=8cm]{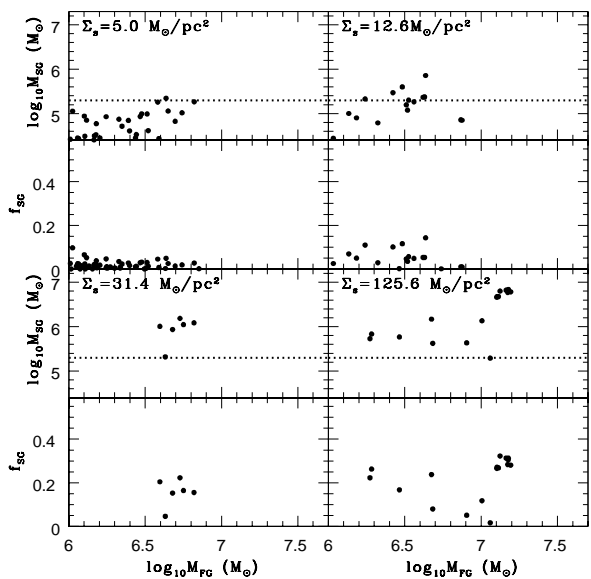}
      \caption{
As in Figure 8 but for
GC candidates in four models with  different $\Sigma_{\rm s}$
(initial mean stellar surface  density):
M13 (upper left),  M14 (upper right),
M1 (lower left), and M15 (lower right).
$f_{\rm g}$ and $f_{\rm b}$ are the same between these four models.
              }
         \label{FigVibStab}
   \end{figure}

   \begin{figure}
   \centering
   \includegraphics[width=8cm]{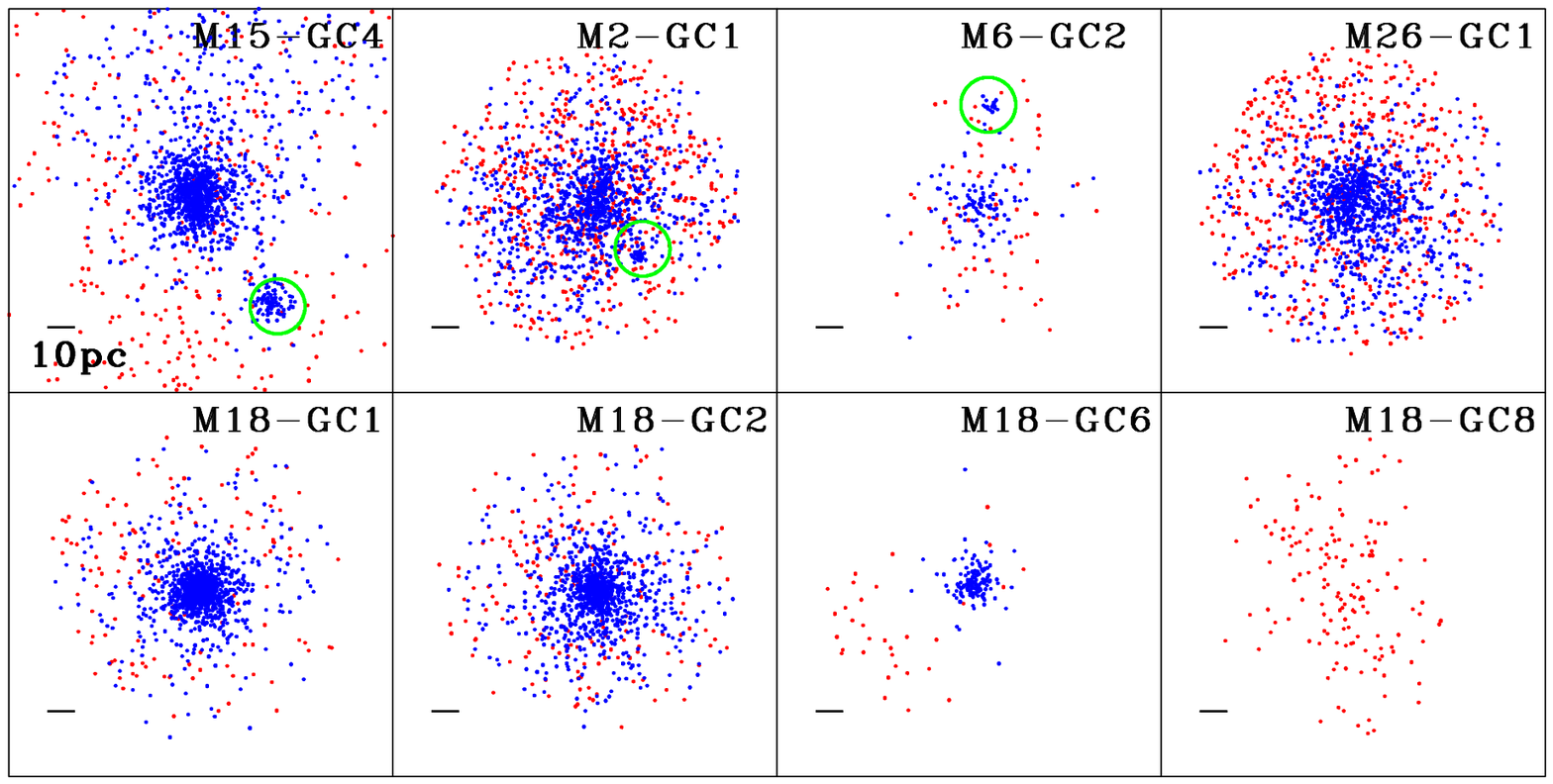}
      \caption{
The distributions of FG (red) and SG (blue) stars projected onto the $x$-$y$ plane for
eight selected GCs. The model number and GC ID are given in the upper right corner for each panel.
For example, `M15-GC4' means that this GC is GC4 in the model M15.
The thick bar in the lower left corner for each panel indicates a size of 10pc.
The three GCs with green circles (M15-GC4, M2-GC1, and M6-GC2) are binary GCs with smaller
companions, and the location of a companion GC is shown by a green circle.
              }
         \label{FigVibStab}
   \end{figure}

\subsection{Parameter dependencies}

\subsubsection{SN and AGB feedback effects}

The following two points are clearly seen in Figure 8. 
First,  SN feedback effects are quite important in controlling both the number of GC candidates 
and the total mass of SG stars in each GC. The number of GCs is lower in the 
model M2 with strong SN feedback effects and higher in
the model M3 without SN feedback effects in comparison with the standard model.
 Only two GCs can be formed in the model with strong SN
feedback effects, and one of the two has an $M_{\rm SG}$ that is too small to be identified as a genuine
GC. The mass fraction of SG stars ($f_{\rm SG}$) is systematically higher in GCs formed
in the model M3 without SN feedback effects.
These results imply that careful modeling for SN feedback effects would be required for
better understanding  the formation efficiency of GCs in dwarfs.
Second, AGB feedback effects can also influence the formation efficiency of genuine GCs
with $M_{\rm SG} \ge 2 \times 10^5  M_{\odot}$. The number of GC candidates 
with $M_{\rm FG} \ge 10^6  M_{\odot}$ (8) in
the model M4 with no AGB feedback effects is larger than that (6) in the standard model, which implies
that thermal and kinetic energy from AGB stars can prevent efficient conversion of AGB ejecta
into SG stars in FG stellar systems.

\subsubsection{Gas mass fraction}

Figure 9 shows how the formation efficiency of genuine GCs with
$M_{\rm SG} \ge 2 \times 10^5  M_{\odot}$ depends on gas mass fractions ($f_{\rm g}$) of
gas-rich dwarfs.
A larger number of GC-like systems  with $M_{\rm GC} \ge 2  \times 10^6  M_{\odot}$ 
can be formed in the model M5 with $f_{\rm g} = 0.67$ and 
$f_{\rm b}=0.07$
and all of them have smaller $M_{\rm SG}$ and diffuse
structures in the SG stellar systems.
A few of them have $M_{\rm SG} \sim 2\times 10^5 M_{\odot}$,
which can be identified as low-mass GCs, 
however, none of the simulated clusters have $M_{\rm SG} >  2\times 10^5 M_{\odot}$.
The model with $f_{\rm b}=0.05$ does not show any GC-like objects
with $M_{\rm SG} \sim 2\times 10^5 M_{\odot}$ (e.g., M11).
The results of M5 and M11 therefore suggest that if $f_{\rm b} \le 0.05$,
then  star clusters formed within dwarf galaxies
are unlikely to evolve into
the present GCs dominated by SG stars. 

The model M6 with  $f_{\rm g}$ and $f_{\rm b}$
higher than those in the model M5 
shows two genuine GCs with $M_{\rm S} \ge 2 \times 10^5  M_{\odot}$,
the number of which is, however, significantly smaller than that derived
in the standard model.  These two results imply that a significantly high $f_{\rm g}$ is 
required for the formation of genuine GCs for lower $f_{\rm b}$. 

The model M7 with a lower $f_{\rm g}$ (=0.4) yet a larger $f_{\rm b}$ shows three genuine GCs,
though $f_{\rm SG}$ is not particularly high for the three. This result suggests that the degree of
self-gravitation in the baryonic component of a dwarf galaxy is a key parameter for
the dwarf to host genuine GCs for a given gas-mass fraction.
The model M8 with a  low $f_{\rm g}$ (0.2) and a higher
$f_{\rm b}$ (0.11) shows four GC-like systems, none of which can be regarded as genuine GCs
owing to their lower $M_{\rm SG}$ and the diffuse structures of their SG stellar systems. 
The combination of $f_{\rm g}$ and $f_{\rm b}$ can change the total gas mass for a given
total dwarf mass (and a given initial mass density of the dwarf). Therefore,
these results imply that there is a threshold total gas mass
(for a given dwarf mass) beyond which the formation of genuine GCs is possible.
Dwarfs are likely to form GCs in their gas disks  only when 
they have a significant amount of cold gas (i.e., only in the early formation phases).

   \begin{figure}
   \centering
   \includegraphics[width=8cm]{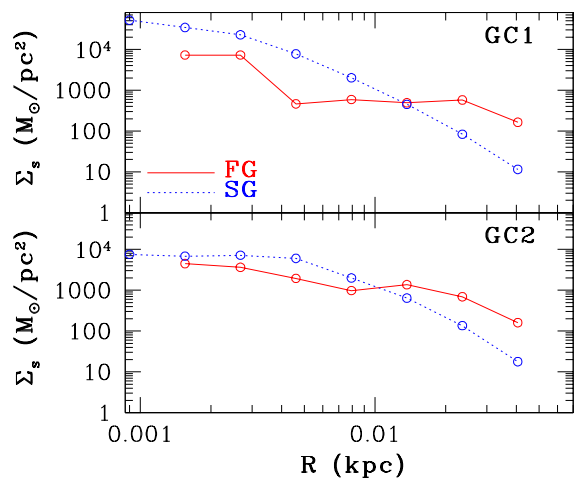}
      \caption{
The same as Figure 7 but for GC1 and GC2 in the model M18. These GCs formed
in a more massive dwarf  are more massive and have higher
central densities than those in the standard model.
              }
         \label{FigVibStab}
   \end{figure}

\subsubsection{Gas surface density}

Figure 10 shows the results of the models M9, M10, M11, and M12
with the same initial mean stellar surface densities
($M_{\rm s}=10^8  M_{\odot}$ and $R_{\rm s}=2.3$ kpc) yet different gas surface
densities ($\Sigma_{\rm g}$). 
Clearly, genuine GCs with $M_{\rm SG} \ge 2 \times 10^5  M_{\odot}$
can be formed in the models M11 and M12
with higher $\Sigma_{\rm g}$ ($\ge 15.6  M_{\odot}$ pc$^{-2}$). 
The mass fraction of SG stars ($f_{\rm g}$) is larger for the models with larger $\Sigma_{\rm g}$.
In the model M10, with lower $\Sigma_{\rm g}$,  only FG stellar systems with low $M_{\rm SG}$ can be 
formed, meaning that long-term dynamical
evolution is unlikely to lead to genuine GCs.
These results mean  that $\Sigma_{\rm g}$ is one of  the key parameters that can determine the formation
processes of GCs.

The results in Figure 10 demonstrate  the presence of diffuse GCs with a smaller fraction
($<0.2$) of SG stars in dwarfs. However,  the observed fraction of SG stars in
GCs with multiple stellar populations is almost always high ($\sim 0.7$; 
Carretta et al. 2010a). This inconsistency implies that such diffuse GCs need
to be destroyed either by their host dwarfs or by the tidal field of the Galaxy
after their accretion onto the Galactic halo. Our future simulations need to
confirm this selective destruction of GCs with smaller fractions of SG stars.
If this is not confirmed, the present model for GC formation
has some serious problems.

\subsubsection{Halo mass}

The following three points are clearly seen in Figure 11. First, there can be a threshold
dark halo mass ($M_{\rm dm}$) beyond which the formation of genuine GCs is possible
for given $f_{\rm g}$ and $f_{\rm b}$. In these four models with $f_{\rm g}=0.8$, $f_{\rm b}=0.11$,
and $R_{\rm g}/R_{\rm s}=2$, only models M1 and M18
with $M_{\rm dm}\ge 2 \times 10^{10}  M_{\odot}$
can host genuine GCs. In the low-mass model M16
 with $M_{\rm dm}=2 \times 10^9  M_{\odot}$,
massive FG systems cannot be formed from a merger of stellar filaments, meaning that AGB ejecta
cannot be efficiently converted into new stars (i.e., SG stars). These results explain why
faint dwarfs in the Local Group do not have any GCs (e.g., van den Bergh 2000).

Second, the number of genuine GCs is larger for the models with larger $M_{\rm dm}$.
It should be noted, however, that the GC formation efficiency ($\epsilon_{\rm gc}=
N_{\rm gc}/M_{\rm dm}$,
where $N_{\rm gc}$ is the total number of GC candidates) is lower in the model
with larger $M_{\rm dm}$: $\epsilon_{\rm gc}=3$GC  per $10^{10}  M_{\odot}$
for $M_{\rm dm}=2 \times 10^{10}  M_{\odot}$
and $\epsilon_{\rm GC}=1.3$GC per $10^{10}  M_{\odot}$
for $M_{\rm dm}=6 \times 10^{10}  M_{\odot}$.
Third,  more massive dwarfs with larger $M_{\rm dm}$  can host GCs.
It is intriguing that the most massive SG stellar systems  in the model M18 
with $M_{\rm dm}=6 \times 10^{10}  M_{\odot}$
is more massive than that in the model M1 with $M_{\rm dm}=2 \times 10^{10}  M_{\odot}$.
Some genuine GCs in the model with $M_{\rm dm} =  6 \times 10^{10}  M_{\odot}$
have larger $f_{\rm SG}$ ($\sim 0.4$) and larger $M_{\rm SG}$ ($\sim 3 \times 10^6  M_{\odot}$),
which could be progenitors of giant GCs in the Galaxy and M31 such as $\omega$ Cen and G1.

These results mean  that (i) more massive GCs are likely to be formed in more massive GC-host dwarfs
and thus (ii) more massive GCs can be more metal-rich because of the mass-metallicity relation
of their host dwarfs (i.e., more metal-rich in more massive dwarfs).
Conclusion (ii)  implies that 
the origin of the  mass-metallicity relation (known as `blue-tilt') of GCs
observed in galaxies (e.g., Strader et al. 2005; Harris et al. 2006) can be  related
{\it not} to chemical evolution within forming GCs but to a trend of more massive GCs
to be formed in more massive dwarfs with more metal-rich gas.

\subsubsection{Low surface brightness versus high surface brightness}

Initial mean stellar surface densities ($\Sigma_{\rm s}$) can also control the formation
processes of GCs.
Figure 12 shows the results of two LSB (M13 and M14) and two HSB models (M1 and M15)
for fixed $M_{\rm dm}$,
$f_{\rm g}$, $f_{\rm b}$, and $R_{\rm g}/R_{\rm s}$.
Although numerous massive clumps with masses larger than $10^6  M_{\odot}$ can be formed
in the LSB model M13, they are dominated by gas and have diffuse stellar distributions
meaning that they cannot become genuine
GCs. In the LSB model M14, only a fraction of GC candidates have 
$M_{\rm SG} \ge 2 \times 10^5  M_{\odot}$ and their SG systems are far less compact
in comparison to those in the standard model. A larger number of
genuine GCs can be formed in the model M15,
some of which are very massive ($M_{\rm GC} \sim 10^7  M_{\odot}$), like
ultra-compact dwarfs.
These results clearly demonstrate that mean stellar surface densities of dwarf galaxies
are important for GC formation and that dwarfs that are formed at higher redshifts and thus likely to
have higher stellar surface densities can host genuine GCs. 

It should be stressed that although the model M17 with $M_{\rm dm}=6 \times 10^9 M_{\odot}$
and  $R_{\rm g}=2.5$ kpc does not show the formation of genuine GCs,
the high-density model M20 with $M_{\rm dm}=6 \times 10^9 M_{\odot}$ and $R_{\rm g}=1.2$ kpc
(i.e., rather high-density dwarf) shows  GC formation.
However, the low-mass yet high-density model M19 with
$M_{\rm dm}=2 \times 10^9 M_{\odot}$ and  smaller $R_{\rm g}$ (=0.7 kpc)
does not show  GC formation. These results combined with those
in Figure 12 imply  that the threshold halo mass
for the formation of genuine GCs with SG stars is around 
$M_{\rm dm}=6 \times 10^9 M_{\odot}$. 
In these low-mass models,  SN feedback effects and more compact dark matter
distributions can cooperate to suppress the formation of FG and SG stars more severely.

\subsubsection{Other minor parameters}

Initial gas temperatures ($T_{\rm g}$) and $Q$ parameters ($Q_{\rm s}$ and $Q_{\rm g}$) can
be different in GC host dwarfs. The models with different $T_{\rm g}$ (M23 and M24)
and different $Q$ (M25, M26, and M27) are therefore investigated so that the dependencies of
the two-stage GC formation processes on these parameters can be understood clearly. The following
interesting results are found. 
Although GC formation processes are not different between M1 and M23 with $T_{\rm g}=10^3$K,
GCs cannot be formed in M25 with relatively high $T_{\rm g}$ ($=10^4$ K). This result implies that
if the ISM of a dwarf
is heated 
by some thermal processes (e.g., energetic stellar winds from massive stars
and reionization effects) prior to GC formation,
then GC formation can be severely suppressed.
The formation of GCs with SG stars can be clearly seen in M25 with low $Q_{\rm g}$ and
M26 with moderately high $Q_{\rm s}$ and $Q_{\rm g}$ (=1.5). However, M27 with high $Q_{\rm s}$
and $Q_{\rm g}$ (=3.0) does not show any GCs with compact SG stellar systems. These results
imply that GC formation is possible only when GC host galaxies are kinematically cold systems.

It is confirmed that the models with lower $\rho_{\rm th, SG}$ do not show GCs with
compact SG stellar systems. For example, the model M28 with $\rho_{\rm th, SG}=1$ atom cm$^{-3}$
shows GC candidates,  but the SG stellar systems of the candidates are so diffuse that
they cannot be regarded as genuine GCs. The diffuse SG system reflects the fact that AGB ejecta 
can be converted into new stars before a strong gaseous condensation can form in the central
region of the FG stellar system through
gaseous dissipation.  On the other hand, in the model M29 with
overly high  $\rho_{\rm th, SG}$ ($1000$ atom cm$^{-3}$),
the formation of SG stars is severely suppressed, meaning that the final SG system cannot
become  particularly compact.
These results for M28 and M29  imply
that physical conditions required for star formation in dense stellar systems
are important for understanding the origin of 
SG stars in GCs.
As discussed in \S 2, there could be great uncertainty  in the mass-metallicity relation of high-$z$ dwarf galaxies. We accordingly investigated
two models in which  [Fe/H]=$-0.7$ and $-2.5$ are adopted yet
other parameters are the same as those of the standard one.
We have  confirmed
that the results do not depend on metallicities:
the degree of clumpiness in dwarf disks 
is slightly different between these models.
The major two-stage GC formation process is not greatly influenced
by the initial metallicities of gas disks of dwarfs.

\subsection{Binary globular cluster formation}

The morphological properties of the simulated GCs in the present study are diverse depending
on the formation processes and their hosts' physical properties. One of the more intriguing results
on GC morphologies is that some GCs have smaller companion GCs {\it with SG stars}.
Figure 13 shows three examples of binary GCs 
(M15-GC4, M2-GC1, and M6-GC2) in which smaller companion GCs can merge
with larger ones {\it only after the formation of compact SG stellar systems.} The smaller 
companion GCs in these three are later captured by larger GCs and finally merge with them
to form single GCs.  These GCs are rarely identified at the final step in each simulation,
because the timescale of GC merging after tidal capture of companion GCs is rather short.
The merging of GCs with SG stars can provide information on the origin of unique characteristics
of some Galactic GCs such as NGC 1851 and M22 (e.g., Carretta et al. 2010b; Bekki \& Yong 2012).

Figure 13 also shows two single GCs (M18-GC1 and M18-GC2) which appear to have global morphologies
similar to other single GCs (such as M26-GC1) yet show higher central mass densities. 
Figure 14, describing the projected radial density profiles in these two GCs, demonstrates that
the projected stellar densities at $R\sim 1$pc are significantly
larger than those for GCs in the standard model. The SG stellar system in M18-GC1 shows
$\Sigma_{\rm s} \sim 4 \times 10^5  M_{\odot}$ pc$^2$ at $R \sim 1$pc. Given that
these two are formed in the massive dwarf model with $M_{\rm dm}=6 \times 10^{10}  M_{\odot}$,
this result implies that GCs with rather high densities are likely to be formed in more 
massive dwarfs. Figure 13 shows that  M18-GC6  and M18-GC8 are dominated by SG and FG stars,
respectively. In the present study, GCs dominated by FG stars are almost always rather diffuse,
and are therefore unlikely to  survive tidal destruction by their host dwarfs and thus are unlikely to be
identified as GCs later.
The animations for the formation of single GC M26-GC1 and binary GC
M16-GC4 are given in Appendix B.

   \begin{figure}
   \centering
   \includegraphics[width=8cm]{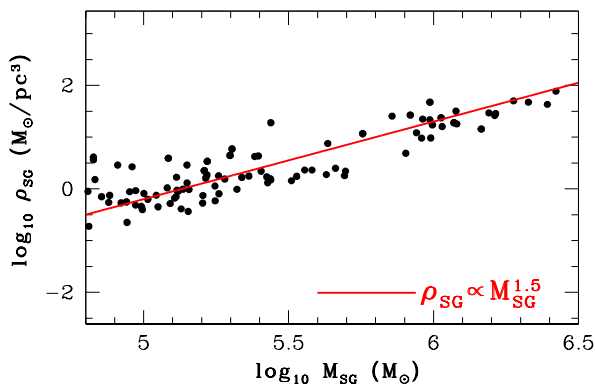}
      \caption{
The plots of GC candidates
with $M_{\rm GC} \ge 2 \times 10^6 {\it M}_{\odot}$
on the $\rho_{\rm SG}-M_{\rm SG}$ plane, where $\rho_{\rm SG}$ is the mean
mass density of SG stars in
an originally massive GC candidate  at its half-mass radius ($R_{\rm eff}$)
of SG stars.
All SG stars within 57pc from the center of a GC candidate are used to estimate
$R_{\rm eff}$ for SG stars of the GC.
These
GC candidates  are formed in the models with $M_{\rm dm}=2 \times 10^{10} {\it M}_{\odot}$.
              }
         \label{FigVibStab}
   \end{figure}

   \begin{figure}
   \centering
   \includegraphics[width=8cm]{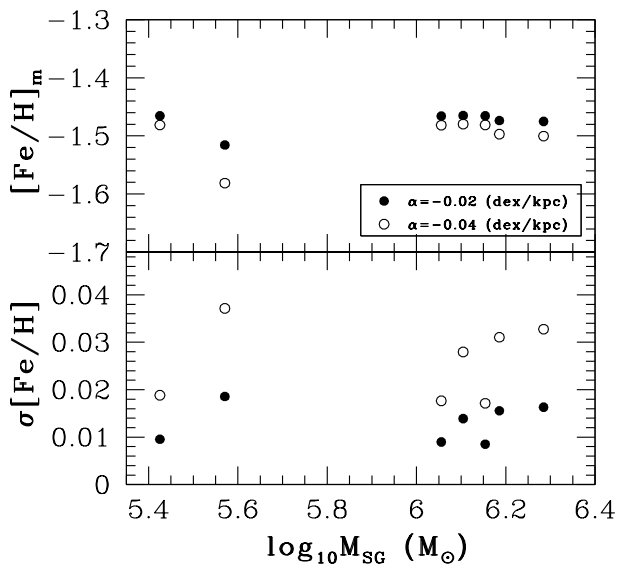}
      \caption{
The dependencies of mean [Fe/H] ([Fe/H]$_{\rm m}$; upper)  and internal [Fe/H] spread
($\sigma$[Fe/H]; lower)
among  GC stars on $M_{\rm SG}$ in the seven GC candidates of the standard model
for $\alpha=-0.02$ (filled circles) and $\alpha=-0.04$ (open circles),
where $\alpha$ is the slope of an initial metallicity gradient (dex kpc$^{-1}$).
              }
         \label{FigVibStab}
   \end{figure}

\subsection{Mass-density scaling relation}

Figure 15 shows a mass-density  ($M_{\rm SG}-\rho_{\rm SG}$) relation  of SG stellar systems in the
simulated GC candidates for the models with $M_{\rm dm}=2 \times 10^{10}  M_{\odot}$. 
Here $\rho_{\rm SG}$ represents the  {\it mean} stellar densities at half-mass
radii ($R_{\rm eff}$)
estimated for stars within 57pc of the simulated GCs. Therefore, $R_{\rm eff}$ 
is significantly large  for the originally massive GCs and accordingly
should not be compared with the observed half-light radii ($\sim 3$ pc) of the present GCs.
The GC candidates have $0.5$ pc $\le R_{\rm eff} \le$ 37.5 pc with a mean $R_{\rm eff}=18.2$ pc,
and some GC candidates have small $M_{\rm SG}$ ($\le 2 \times 10^5  M_{\odot}$).
The present simulations with a spatial resolution of at most $\sim 1$pc can only  allow us to
discuss a possible relation between {\it  mean} mass densities and masses of GCs.
Also, it is reasonable to investigate the mass-density relation for SG stellar systems to provide
some clues to the origin of the observed mass-density relation of the Galactic GCs,
because the majority ($\sim 70$\%) of the present stellar populations
of the GCs  are FG stars (e.g., Carretta et al. 2010a).

Figure 15 shows that $\rho_{\rm SG}$ is higher for larger $M_{\rm SG}$ for GCs and
the  $M_{\rm SG}-\rho_{\rm SG}$ relation can be well approximated by 
$\rho_{\rm SG} \propto M_{\rm SG}^{1.5}$. The physical reason for this relation can be described
as follows. In the present models,  
AGB and SN feedback effects can prevent gas from being converted into new stars, in particular
for SG stars. These feedback effects can be more severe for forming GCs with lower masses owing
to shallower gravitational potentials. As a result of this, a smaller  amount of AGB ejecta
can be converted into SG stars within FG stellar systems for forming GCs with lower masses.
Thus, less  massive GCs with smaller  $M_{\rm SG}$ can have lower $\rho_{\rm SG}$.

The derived $M_{\rm SG}-\rho_{\rm SG}$ relation can be compared with the observationally expected
one by using the $L-\sigma$ (i.e., luminosity-dispersion) relation derived by Djorgovski (1993). 
The Galactic GCs have the $L-\sigma$ relation:
\begin{equation}
L \sim \sigma^{\beta},
\end{equation}
where $\beta = 1.65$ ($\pm 0.15$). By using the virial theorem ($M \sim R \sigma^2$) 
and assuming a
universal $M/L$,
the above equation can be rewritten as follows:
\begin{equation}
\rho \sim M/R^3 \sim M^{1.61}.
\end{equation}
This mass-density relation (or size-mass relation of $R \propto M^{-0.21}$)
is very similar to the simulated one ($\rho_{\rm SG} \propto M_{\rm SG}^{1.5}$),
which implies that the origin of the observed scaling relations of GC (e.g., Djorgovski 1993)
could be closely related to the formation processes of SG stars in  FG stellar systems.

\section{Discussion}

\subsection{Formation of `first generation' systems in gas-rich, baryon-rich
dwarfs at high z}

Carretta et al. (2010a) have revealed that
the fraction of stars (SG) that were formed from
gas chemically polluted by earlier generations  of stars (i.e., FG stars, e.g., FRMS and AGB stars) in GCs with internal variation of 
light elements are quite large (70\%).
The initial total  masses of FG systems
required for explaining the observed large
SG fractions are  at least ten times more massive than the present
masses of GCs (e.g., D08 and B11). Therefore,  previous theoretical models
of GC formation (D08, B11) adopted an assumption of initially
very massive stellar systems and thereby discussed
how SG stars were formed and evolved. 
The required mass for FG systems appears to be uncomfortably large 
(`mass budget problem'), and 
Larsen et al. (2012) suggested that GC formation scenarios
based on the very massive FG systems
cannot explain the observed small fraction of metal-poor stars
and high GC specific frequency in the Fornax dwarf galaxy self-consistently.

There are three ways (scenarios) to overcome this mass budget problem. The first is
to adopt a contrived very top-heavy IMF for star formation of FG systems
(e.g., Prantzos \& Charbonnel 2006).
 This can significantly reduce the required initial masses of FGs
because of a much larger fraction of AGB stars or massive stars,
and such a top-heavy IMF is suggested to be possible in low-metallicity
environments (e.g., Marks et al. 2012). However, GCs with such a top-heavy
IMF are highly likely to disintegrate quickly (Bekki \& Norris 2006),
and thus rapid accumulation of gas in the FG systems could be less
likely. The possibility of gas accumulation in rapidly disintegrating
FG systems with a top-heavy IMF needs to be investigated numerically
in this first scenario.

The second scenario assumes that gas for SG system formation originates
not from massive star clusters but from field stars in the central
regions of dwarf galaxies. In this scenario,  gaseous ejecta from central field
stars and pristine ISM can be converted into new stars in the nuclear
regions of dwarfs meaning that SG systems can form and can be identified
as nuclear GCs (e.g., Bekki 2006; Bekki et al. 2007; Maxwell et al. 2014).
The nucleated dwarfs can be transformed into naked nuclei (i.e., GCs)
during their accretion onto the Galaxy where their stellar envelopes
are completely stripped.
There is no budget problem in this scenario, because GC host galaxy
itself can provide a large amount of gas for the SG system formation.

The third scenario involves  adopting an assumption that GCs with multiple stellar 
populations are composed of two (or more) populations that were 
formed at different epochs.
It could be possible that
the origin of the observed abundance spreads is not due to two different
major star formation events with the earlier star formation event
influencing
the chemical abundances of stars formed in the later one.
Bastian et al. (2013) indeed proposed that gas ejecta from
massive interacting binaries in a  forming GC
can chemically pollute some fraction of  low-mass pre-main
sequence stars, and this can cause stellar populations
with different chemical abundances.

If GC progenitors are really as massive as $10^6 - 10^7  M_{\odot}$,
then the physical mechanisms for their formation need to be understood.
The present study suggests that massive gas clumps in  gas-rich dwarfs
are the formation sites of FG systems.
It also suggests that 
such massive FG systems are more likely to be formed in high-z
dwarfs,  because they are very gas-rich  and have
mass densities owing to earlier virialization,
meaning that local gravitational instability can lead to
the formation of massive clumps.

Furthermore, the simulated FG systems in the present study
do not look like spherical systems when SG stars begin to form.
FG systems form from mergers of smaller stellar clumps and filaments
originating from different regions in a dwarf,
meaning that they cannot be fully developed and dynamically relaxed
before their AGB stars eject gas. 
This suggests that the fully developed spherical stellar system
adopted in previous theoretical studies of GC formation
could be oversimplified or unrealistic.
The gas accumulation process within FG systems and secondary SF 
from gas could be significantly different between
fully developed FG systems and growing FG systems.
Therfore,  future theoretical models of GC formation
need to consider hierarchical growth of FG stellar systems through
mergers of small clumps and filaments in order to more accurately
predict chemical abundances of SG stars formed from gas ejecta
from FG stars.

\subsection{The dilution of AGB ejecta by pristine gas}

Although most of the previous  chemical evolution  models have shown that dilution of AGB ejecta
from FG stars by pristine gas is required to explain the observed star-to-star variation of
chemical abundances in GCs (e.g., Bekki et al. 2007; D'Ercole et al. 2010),
it remains theoretically unclear where such pristine gas can come from to mix with AGB ejecta
in forming GCs
(D'Ercole et al. 2011 for a recent summary on this issue).
Original  GCs can obtain pristine gas by interaction with nearby giant molecular clouds
(Bekki \& Mackey 2009) via Bondi-accretion of ISM
(Pflamm-Altenburg \& Kroupa 2009; Conroy \& Spergel 2011) and  gas accretion from unevolved stars within their  FG stellar systems
(`self-dilution'; Gratton \& Carretta 2010). D'Ercole et al. (2011) concluded that 
theoretical models based on Bondi-accretion (Conroy \& Spergel 2011)
and self-dilution (Gratton \& Carretta 2010)
cannot explain the observed presence of stars with large He abundances ($Y>0.35$) in some GCs.
Since these theoretical studies  are based on  simple analytical models,
it is unclear whether the proposed dilution processes can really occur 
at the epoch of  GC formation.

The present study has shown that pristine gas can be captured by  FG stellar systems
{\it after the ejection of gas from AGB stars in the FG systems}
and that it is initially located in the low-density regions of local  gaseous 
clumps/filaments 
(see Fig. 4).
ISM  that is not expelled from  dwarf galaxies by SN explosion
and  is not converted into new stars owing to its low-density can be captured
later by FG stellar systems, meaning that it can be regarded as `pristine'. 
Therefore, the pristine gas required for the formation of SG stars 
in the two-stage GC formation originates from clumpy/filamentary gaseous structures developed from
local gravitational instability in gas disks of dwarf galaxies.
As shown in the present study,
the pristine gas comes originally  from different local regions within its host dwarf.
It is therefore possible that if dwarf galaxies  have large radial gradients of chemical abundances,
then the pristine gas for GC formation can also have a large dispersion in chemical abundances.
We discuss this point below.

Chemical influences of {\it field} massive and AGB stars (i.e., those
already existing before
the formation of FG stars in GCs)  on local gaseous clumps/filaments
are ignored in the present study. Therefore, it is not clear whether 
gas in local clumps and filaments that can be later captured by
FG stellar systems can be really pristine 
(i.e., the gaseous chemical abundances  similar to those of FG stars) at the formation
epoch of SG stars. Given the formation timescale of at least $\sim 200$ Myr in the
two-stage GC formation model,  the field massive and AGB stars could possibly
increase the dispersion in chemical abundances through mixing of their ejecta with
ISM.
This point should be properly addressed by our future studies based on
full chemodynamical simulations with self-consistent inclusion of both SN explosions
and AGB winds.

\subsection{High-z gas-rich dwarfs as a preferred formation site of GCs
with multiple stellar populations}

The present study suggests that gas-rich dwarfs virialized at high z
are more likely to form GCs with multiple stellar populations,
because massive stellar and gaseous clumps/filaments, which can finally
become FG stellar systems, can be developed through local gravitational
instability. Massive FG stellar systems 
cannot only retain gas ejected from AGB stars (D08; B11)
but also capture ISM that is mixed with the AGB  ejecta for
secondary star formation (e.g., Pflamm-Altenburg \& Kroupa 2009).
Therefore, the initial total masses of FG systems can determine
whether the final dense stellar systems  can become bona-fide GCs with
internal variations of light elements (e.g., NGC 1851; Yong et al. 2009; 2015)
or those without
(e.g., Ruprecht 106; Villanova et al. 2013).

However, this scenario cannot simply explain the existence of
the Galactic open cluster NGC 6791 with [Fe/H]=0.4 and internal variation
of light elements (Geisler et al. 2012). Its high  metallicity,
 relatively younger age (4 Gyr), and slightly supersolar [O/Fe]
(e.g., Carroro et al. 2006) strongly suggests that this cluster
was formed in the central region of the Galaxy,
where gas mass fraction should have been low at the cluster
formation.
Therefore the physical properties of the low-mass cluster
NGC 6791 imply that gas ejected from FG stars is retained
by an unknown mechanism (other than massive FG stellar systems).
 It could be possible that
the cluster comes from the inner bulge, where the deep gravitational well
of the bulge might have enabled the cluster to retain the AGB ejecta.
The origin of this cluster still remains unclear.

\subsection{Possible [Fe/H] spread in GC stars}

In  the present two-stage GC formation scenario,
gas from which FG stars can form needs to originate from different local clumps and filaments
so that the final GCs composed of FG and SG stars
can become as massive as  $2 \times 10^6  M_{\odot}$.
Therefore, it would be inevitable that GCs can have internal [Fe/H] spread,
if their host dwarf galaxies have chemical abundance spreads in their gas disks.
We investigate  quantitatively a possible internal [Fe/H] spread for each of the simulated
GCs in the standard model as follows.
We first allocate  metallicity to  each gaseous particle in the disk of the dwarf 
according to its initial position: at  $r$ = $R$
where $r$ ($R$) is the projected distance (in units of kpc)
from the center of the disk, the metallicity of the gas particle is given as:
\begin{equation}
{\rm [Fe/H]}_{\rm r=R} = {\rm [Fe/H]}_{\rm d, r=0} + \alpha \times {\rm 
R}. \;
\end{equation}
We then search for the initial locations (within the gas disk at $T=0$) of FG and SG stars in a GC
and thereby investigate initial [Fe/H] of all stars in the GC by using the above equation.
We consider  that the slope $\alpha$ is a free parameter and
that ${\rm [Fe/H]}_{\rm d, r=0}$ is set to be $-1.45$ for this model.
We investigate the model with $\alpha=-0.02$ and $-0.04$ (dex kpc$^{-1}$),
the observed  value  for the Galaxy
(e.g., Andrievsky et al. 2004).

As shown in Figure 16,  internal [Fe/H] spread ($\sigma$[Fe/H]) for each of the seven GC candidates
in the standard model depends on $\alpha$ such that
$\sigma$[Fe/H] is larger for steeper initial metallicity gradients.
This result is essentially consistent with our previous simulations (without the formation
of SG stars) on the origin of abundance spreads in heavy elements for the Galactic GCs (Bekki 2012).
For example,  $\sigma$[Fe/H] can be  $\sim 0.02$ dex for $\alpha = -0.02$
and $\sim 0.04$ dex for $\alpha = -0.04$ in the standard model.
The mean [Fe/H] ([Fe/H]$_{\rm m}$) does not depend strongly  on $\alpha$ in this model.
These results imply that the observed $\sigma$[Fe/H] in GCs can provide some constraints
on $\alpha$ in their host dwarfs at the epoch of GC formation. 
The present study has shown that GC formation can be possible when their host dwarfs
are relatively  gas-rich and thus in the early formation phases. It is possible that $\alpha$
is rather small in the early phase of dwarf formation  owing to less advanced chemical enrichment.
The present study suggests that if the formation of GCs  with SG stars is possible mostly
in the early formation phases of dwarfs, then $\sigma$[Fe/H] should be small.

These results on $\sigma$[Fe/H] could provide a clue to the origin of recent observations
on possible Ca abundance spreads in some Galactic GCs (e.g., Lee et al. 2009).
In the present two-stage GC formation model, however, 
$\sigma$[Fe/H] can be at most $\sim 0.1$ dex for $\alpha > -0.1$ (dex kpc$^{-1}$).
Therefore, the present study cannot explain the observed larger spreads in heavy elements
in some massive Galactic GCs: 1.5 dex in  $\omega$ Cen (Norris \& Da Costa 1995),
0.2 dex in M54 (e.g, Bellazzini et al. 2008), 0.2 dex for Ca in NGC 2419 
(e.g., Cohen \& Kirby 2012), 0.15 dex in M22 (e.g., Da Costa et al. 2009;
Marino  et al. 2011), and 0.3 dex in Terzan 5 (Ferraro et al. 2009).
Such large spreads in these GCs 
cannot be simply explained by initial abundance spreads of heavy elements
in the gas disks of their host galaxies.
Other physical mechanisms such as GC formation in the central regions
of galaxies where continuous gas supply is possible 
would be required for explaining these GCs with large abundance spreads of heavy elements.

It should be noted here that possible internal metallicity spreads among
FG stars (i.e., those generated during star formation in GC-forming
molecular clouds) are not considered in the above discussion:
$\sigma$[Fe/H] in Fig. 16 can be a minimum possible spread.
Our recent hydrodynamical simulations  of GC formation
within fractal molecular clouds
were used to investigate [Fe/H] spreads among FG stars
in simulated GCs and found that the typical spread is  less than 0.05 dex
(Bekki 2017).
This small spread of 0.05 dex is comparable to the derived initial [Fe/H] spread
of GC-forming molecular clouds in Fig. 16.

\subsection{Comments on the latest observations of star clusters
in the Magellanic Clouds}

Intermediate-age and young Massive star clusters in the LMC
are observed to have  extended main sequence turn-offs (eMSTOs).
Whether multiple generations  of
stars (i.e., internal age spread)
or stellar rotation can be responsible for the eMSTO phenomena
remains unclear (Mackey \& Brodby Nielsen 2007; Bastian et al. 2013;
Goudfrooij et al. 2014; Cabrera-Ziri et al. 2016;  Li et al. 2016).
Recently, FB17 discovered young stellar objects
(YSOs) within the central regions (less than the central 1pc)
of seven LMC clusters with ages ranging from
100 Myr to 1 Gyr. Given that the ages of YSOs are less than 1 Myr,
their results confirm that there can be very young stellar populations
in some of the LMC clusters. 
It has recently been confirmed that the YSOs cannot be AGB stars,
because no AGB candidates were found in the color magnitude diagrams of the
stars close to YSOs for the seven LMC clusters
(Bekki et al. 2017).

Furthermore, Milone et al. (2017)
found that the presence of younger stellar populations
and stellar rotation is necessary to reproduce the observed
color magnitude relation of the LMC younger cluster, NGC 1866.
These latest observations demonstrate that some of the LMC clusters have
multiple generations of stars, though the origin of the multiple stellar populations
with different ages is not clear.

FB17 found that
star clusters with YSOs are not necessarily surrounded by high-density cold
gas and thus suggested that new star formation from accreted gas from ISM
cannot be the major mechanism for the multiple generations of stars observed.
The present study demonstrates that secondary star formation is possible
from pure AGB ejecta from forming GCs
 in gas-rich dwarf galaxies. However, some of the star clusters
with YSOs in FB17 are less massive ($<10^5 M_{\odot}$), meaning that they cannot retain
AGB ejecta. Therefore, the physical origin of secondary star formation in
star clusters with YSOs remains unclear. One possibility is that these
star clusters interacted with nearby massive molecular clouds to `snatch' gas
from the clouds for secondary star formation (Bekki \& Mackey 2009).

\section{Conclusions}

We have investigated how GCs with multiple stellar populations can be formed from star-forming
gas clouds
by adopting a novel method to resolve the two-stage GC formation
in  numerical  simulations
of gas-rich dwarf disk galaxies.
A principle assumption in the present simulations 
is that SG stars can  be formed  from gas ejected from FG AGB
stars in forming GCs.
The present numerical simulations are different from previous ones in that 
they enable us to investigate, for the first time,
both the formation process of FG stars and that of SG stars
from gaseous ejecta of FG stars in a self-consistent manner. 
The principal results are summarized as follows. \\

\begin{enumerate}
\item
The formation process of GCs is basically two-stage. 
First,  new stars (i.e., FG stars) can form efficiently from the high-density regions
of gaseous clumps and filaments
developed through local gravitational instability in the gas disks of dwarfs.
The local  stellar clumps (or sub-clusters) and filaments merge with one another to form 
massive  diffuse stellar systems
with masses larger than $10^6  M_{\odot}$.
Then,  gas ejected from AGB stars evolved from these FG stars 
can finally sink into the potential well of the FG stellar system and
can subsequently be
converted into new stars corresponding to SG stars.  
The formation timescale for GCs with FG and SG stars is about 200 Myr in the present
two-stage GC formation process.
Although the formation of SG stars in
FG stellar systems has already been suggested and  investigated by previous studies
(e.g., D08, B11),
the present study has shown, for the first time, the entire two-stage formation process
of GCs with FG and SG stars. 
It should be noted, however, that SG stars are assumed to be formed from 
AGB ejecta only if local gas density exceeds a threshold gas density in the
present study. This assumption could be over-simplified for star formation
in existing dense stellar systems.

\item
Gas in the low-density regions of local clumps and filaments does not form stars, and consequently
can be captured later by nearby FG stellar systems in forming GCs. Such gas can therefore be used
as `pristine gas' for mixing of AGB ejecta to form SG stars in the central regions of 
FG stellar systems. The present simulations suggest that  AGB ejecta can be diluted by
ISM that 
(i) does not participate in the formation of FG stars,
(ii) is not expelled from GC host galaxies through SN explosions,
and (iii) is close to forming FGs.
AGB ejecta both from FG  stars in a GC and from field stars
(that do not finally become member stars of the GC)
can contribute  to the formation of SG stars in the GC.
More sophisticated high-resolution simulations will be necessary to confirm
that such ISM can really have chemical abundances  similar 
to those of FG stars. 

\item
SG stellar systems in simulated GCs can have more compact spatial distributions
and larger rotational amplitudes in comparison with the FG stellar systems.
These structural and kinematical differences between FG and SG stellar systems are due largely
to dissipative formation of SG stars in FG stellar systems. The mass fractions and spatial 
distributions of SG stars within GCs are diverse, and some diffuse GCs can have no/few SG stars.
FG stellar systems can have rotational kinematics in some massive GCs owing to mergers
of sub-clusters at their formation phases. 

\item
SG stellar systems with larger total masses ($M_{\rm SG}$) are likely to 
have higher mass densities ($\rho_{\rm SG}$), and the mass-density relation
can be roughly described as $\rho_{\rm SG} \propto M_{\rm SG}^{1.5}$.  
More massive GCs are likely to be formed in more massive dwarfs. Given the observed
mass-metallicity relation of dwarfs (i.e., more massive dwarfs being more metal-rich),
this result suggests that more massive GCs are likely
to be more metal-rich. Thus the origin of the observed possible mass-metallicity
relation in metal-poor GCs (`blue tilt') could  be related 
to a trend for more massive GCs to be formed in more massive dwarfs. 

\item
It is inevitable that massive GCs can show internal [Fe/H] spreads in the present two-stage GC
formation because FG stars in such GCs need to 
form from gas clumps and filaments initially located in
different regions of dwarfs.  The [Fe/H] spreads ($\sigma$[Fe/H]) depend on radial metallicity
gradients ($\alpha$ dex kpc$^{-1}$) in disks of GC host dwarfs
such that $\sigma$[Fe/H] is larger for steeper gradients.
The possible $\sigma$[Fe/H] is, however,  as small as 0.04 dex for $\alpha=-0.04$ for GCs formed
in dwarfs with $M_{\rm dm}=2 \times 10^{10}  M_{\odot}$. 
It should be noted here that our simulations do not include self-enrichment
by SNe in forming GCs. Therefore, the above $\sigma$[Fe/H] is a minimum possible
value.

\item
Binary GC formation via the capture of a smaller GC by a larger one is clearly
seen in the present hydrodynamical simulations in which GC formation from ISM and AGB ejecta
is self-consistently investigated. 
However,  the timescale of a binary GC status is short
($ \sim 10^8$ yr) because of the subsequent rapid GC-GC merger in the halo of FG stellar systems.
The merger of GCs {\it with compact SG stellar systems}
(i.e., merging after SG formation)  is demonstrated to be possible, which provides
a clue to the origin of some Galactic GCs with unique characteristics
of multiple stellar populations (e.g., NGC 1851 and M22).

\item
There is  a threshold mass of GC host galaxies ($M_{\rm h,min}$)
beyond which the formation of GCs with FG and SG 
is possible. The possible $M_{\rm h,min}$ including dark matter halos of dwarf disks 
can be $[5-23] \times 10^{9}  M_{\odot}$. 
Below this $M_{\rm h,min}$, diffuse GC-like objects can be
formed in disks, but the total masses of SG stars in these objects are much less than
$\sim 10^5  M_{\odot}$.  They are highly likely to be easily  destroyed by the tidal
fields of their host galaxies and thus unlikely to be the present GCs with multiple stellar
populations.  The baryonic fraction ($f_{\rm b}$)
and gas mass fractions ($f_{\rm g}$) of dwarfs' disks are also 
key parameters that determine whether GCs can have SG stars.
Galaxies with higher $f_{\rm b}$ are more likely to form GCs.

\item
The required  high $f_{\rm b}$ and $f_{\rm g}$ 
imply that dwarf galaxies at high $z$ ($>5$)
could be the preferred formation sites of GCs with multiple stellar population.
The required $f_{\rm b}$ and $f_{\rm g}$ also suggest
that the physical properties of GCs predicted in previous
cosmological simulations (e.g., Bekki et al. 2008)
would not be particularly robust because the simulations did not properly
consider these threshold values.
The present constrained simulations, however, did not allow us
to investigate whether such high 
$f_{\rm b}$ and $f_{\rm g}$  can really be achieved within
high-$z$ dwarfs in a cosmological context.

\item
 Both AGB and SN feedback effects are important for the formation of GCs with FG and SGs stars
in the simulated two-stage GC formation process,
because they can suppress the formation of SG stars in the central regions of the FG stellar
systems.  It is, however, unclear how AGB stars and SNe in FG systems
can influence the chemical abundances of gas from which SG stars can form in the present study.
Thus, although the present study has shown that the formation of `multi-generations' of stars
is possible in dwarf galaxies,
it remains unclear whether the simulated 
different generations  of stars in GCs  have chemical abundances consistent
with the observed ones in GCs.

\item
Previous theoretical models of GC formation
with multiple stellar populations
assumed  fully developed massive  stellar systems
in order to discuss chemical and dynamical properties of 
different stellar populations
(D08, B11,  Bastian et al. 2013).
 The present results suggest that such an assumption
could be unrealistic. Given that FG stellar systems form through
mergers of smaller clumps and filament-like structures,
accretion of AGB ejecta and mixing of the ejecta with pristine ISM
in forming FG stellar systems
can be significantly more complicated than what is described in
previous models (D08 and B11).
Furthermore,  AGB ejecta  that can be used for the formation
of SG stars in a  GC can originate not only from the 
stars of the FG system
but also from those which do not finally become the members of the FG system.
Although this alleviates the mass budget problem,
the original masses of FG systems should still be significantly more 
massive than the present GCs.
Future chemical evolution models of GCs based on the formation of SG systems
from gas ejected from FG stars thus
need to incorporate the above formation process of FG systems.

   \end{enumerate}

The present study is a first step toward comprehensive modeling of the sequential formation
processes of FG and SG stars of GCs in galaxies. The numerical study presented here  would be  rather idealized
and less realistic in terms of modeling AGB ejecta and star formation 
in dense stellar systems.
Furthermore, the present model is incapable of investigating  the observed
anti-correlations between light elements (e.g., C, N, and O) 
of GCs with multiple stellar populations. We therefore plan to investigate the origin of the anti-correlations in the multiple stellar populations
of GCs
by using fully self-consistent chemodynamical models with a more sophisticated model for the evolution
of AGB ejecta.
We also need to perform hydrodynamical simulations
of GC formation  for other self-enrichment scenarios (e.g., FRMS,
MIB) in order to assess the validity of each of these scenarios.

%

\begin{acknowledgements}

I am  grateful to the anonymous referee for constructive and
useful comments.
The preliminary results of the present simulations have been
reported in the Special Session 1 of IAU XXVIII meeting
(`Origin and Complexity of Massive Star clusters') held
in Beijing, China, on August 20-24, 2012.
I (Kenji Bekki) am grateful to participants of the meeting who
discussed the results with me during the meeting.
Numerical simulations  reported here were carried out on
the three GPU clusters,  Pleiades, Fornax,
and gSTAR, kindly made available by International
Center for radio astronomy research (ICRAR) at  The University of Western Australia,
iVEC,  and the Center for Astrophysics and Supercomputing
in the Swinburne University, respectively.

\end{acknowledgements}

%
%

\begin{appendix} 

\section{Two animations}

We here present two animations of typical GC formation in the present numerical simulations
in order to help readers to understand  the key formation processes of GCs with
FG and SG stars.

\subsection{The two-stage GC formation process}

The animation file gc1.avi clearly describes the two-stage GC formation process
for one of the GCs (GC1, shown in Figure 13 as M26-GC1)
formed in the model M26.
In this animation, the left and right flames describe the time evolution
of particle distributions  projected onto the $x$-$y$ plane  for
ISM (blue),  FG stars formed from ISM (magenta),
AGB ejecta (green), and SG stars formed from AGB ejecta of FG
stars (cyan) for the large-scale (left) and small-scale (right)  views. The time indicated
in the upper part of each frame is given in units of Myr and the white bar shows a physical scale
(1 kpc in the left and 10 pc in the right).
The center of each frame is coincident with the center of mass of the simulated GC.
This GC is the most massive one with $1.9 \times 10^7  M_{\odot}$ for the initial FG stars
and $2.2 \times 10^6  M_{\odot}$ for the SG stars. Although, the total masses of FG and SG
stars can decrease significantly from these larger masses owing to later dynamical evolution,
this GC candidate is highly likely to become one of the present massive GCs.

\subsection{Formation and evolution of binary GCs}

The animation file gc2.avi  describes formation and evolution of a binary GC (GC4,
shown in Figure 13 as M15-GC4)
formed in model M15. It is not easy to identify a binary GC in a simulation
because simulation data is output at every $1.4 \times 10^8$ yr
for most models (at every $56$ Myr for the standard model)
in the present study.
This binary GC is one of the GCs which happened to be identified during the GC identification process
in the present study.
In this animation, the left and right frames describe the time evolution
of particle distributions  for
ISM (blue),  FG stars formed from gas (magenta),
AGB ejecta (green), and SG stars formed from AGB ejecta of FG
stars (cyan) for the $x$-$y$ (left) and $x$-$z$  projections (right). The time indicated
in the upper part of each frame is given in units of Myr and the white bar shows
a physical size of 10pc.
The center of each frame is coincident with the center of mass of the simulated GC.
This GC is also massive  with $4.7 \times 10^6  M_{\odot}$ for FG stars
and $1.5 \times 10^6  M_{\odot}$ for SG stars.
It is clear that a smaller GC with SG stars is tidally captured by a larger GC with SG stars.
Thus this is a merging between GCs with SG stars (i.e., after SG formation in FG stellar systems).

\end{appendix}

\end{document}